\documentclass[11pt]{article}
\usepackage[margin=1in]{geometry}                
\usepackage{graphicx}
\usepackage[numbers]{natbib}
\usepackage{amssymb,amsmath,bbm,float,dsfont}
\usepackage{epstopdf}
\usepackage{numprint}
\usepackage{xcolor}
\usepackage{subcaption}
\usepackage[font=small,skip=10pt]{caption}

\graphicspath{{./Fig/}}

\usepackage{tikz}
\usetikzlibrary{decorations.shapes}
\usetikzlibrary{decorations.pathmorphing}
\usetikzlibrary{shapes.geometric,arrows}
\usetikzlibrary{fit,shapes}
\tikzstyle{var}=[ellipse,thick,draw=black,minimum size=1.2cm]
\usetikzlibrary{snakes}

\tikzstyle{decision} = [ diamond, draw, fill=blue!20, text width=4.5em, text badly centered, node distance=3cm, inner sep-0pt]  
\tikzstyle{block} = [ rectangle, draw, fill=blue!20, text width=5em, text badly centered, rounded corners, minimum height=4em]  
\tikzstyle{line} = [ draw, -latex']  
\tikzstyle{terminator} = [ draw, ellipse, node distance=3cm, thick,minimum height=3em]  

\newtheorem{assum}{Assumption}

\newtheorem{proposition}{Proposition}
\title{Assessing model prediction performance for the expected cumulative number of recurrent events}
\author{Olivier Bouaziz}
\date{
	Universit\'e Paris Cit\'e, CNRS, MAP5 UMR 8145, F-75006 Paris, France \\}

\begin{document}
\maketitle

\begin{abstract}
	In a recurrent event setting, we introduce a new score designed to evaluate the prediction ability, for a given model, of the expected cumulative number of recurrent events. This score allows to take into account the individual history of a patient through its external covariates and can be seen as an extension of the Brier Score for single time to event data but works for recurrent events with or without a terminal event. Theoretical results are provided that show that under standard assumptions in a recurrent event context, our score can be asymptotically decomposed as the sum of the theoretical mean squared error between the model and the true expected cumulative number of recurrent events and an inseparability term that does not depend on the model. This decomposition is further illustrated on simulations studies. It is also shown that this score should be used in comparison with a reference model, such as a nonparametric estimator that does not include the covariates. 
Finally, the score is applied for the prediction of hospitalisations on a dataset of patients suffering from atrial fibrillation and a comparison of the prediction performances of different models, such as the Cox model or the Aalen Model, is investigated.
	
\end{abstract}


\section{Introduction}

Recurrent event data are often encountered in follow-up studies. They can be seen as a generalisation of the standard time to event data, where individuals may experience the same event repeatedly over time. Typical examples may include HIV studies where patients can experience repeated opportunistic infections, remission data from Leukemia patients who can experience multiple relapses, repeated seizures for epileptic patients, or hospitalisation data where the events of interest are the hospitalisations. 
In those studies, the focus might be on assessing the effect of covariates on the risk of recurrences or on predicting the future recurrences. The first model to deal with recurrent event data was the Andersen-Gill model~\cite{andersen1982cox} which was further extended by~\cite{lin2000semiparametric} to account for possibly dependent jumps of the recurrent event process. Further models were developed such as in~\cite{prentice1981regression}, ~\cite{cook1997marginal}, ~\cite{ghosh2002marginal}, ~\cite{ghosh2003semiparametric} or~\cite{andersen2019modeling} where the last four papers incorporate the presence of a terminal event in the estimation procedure, or using random effects such as in~\cite{hougaard2000analysis}, ~\cite{liu2004shared}, or~\cite{rondeau07}. In particular, in ~\cite{cook1997marginal}, ~\cite{ghosh2002marginal}, ~\cite{ghosh2003semiparametric}, ~\cite{andersen2019modeling} the authors focused on the estimation of the expected cumulative number of recurrent events. This is a marginal quantity that computes the expectation of the number of experienced events of an individual before any time point. This quantity is particularly interesting as it summarises the evolution of the recurrent event process with time. In the presence of a terminal event, it also includes the fact that when the terminal event occurs the patient can no longer experience any further recurrent events.

In some studies the focus is more on the predictiveness ability of a model rather than on the interpretation of the covariates effects. This is the case when clinicians aim at predicting the future repeated events in order to offer the best medical care. Being able to predict the future recurrences of any patients on a short time period also allows to predict the future burden of the disease over the patient's life. Moreover, a predictive model can be an important tool for making medical decisions but also for communicating with the patients about the future course of his/her disease. For instance, in~\cite{schroder2019recurrent} the authors studied patients with atrial fibrillation, a well known cardiac disease, in an attempt to predict the future hospitalisations of patients due to their disease. Since the patients suffering from this disease are usually old (the median age in the study was $63$ years) and since atrial fibrillation can be a severe disease in some cases, those patients were also at risk of death. Several covariates were collected and a prediction of the expected cumulative number of recurrent events over time was performed using a Cox model with dependence on prior counts. 

While such models are certainly of interest for clinicians, it is important to propose relevant diagnosis tools that can evaluate the prediction performance of the proposed model. There already exists several indicators for prediction performances in the standard context of time to event data with only one event per individual. The Brier score was developed in~\cite{graf1999assessment} and in~\cite{gerds2006consistent}, which basically is a score for computing the mean squared error of the time to event in the presence of censoring. This score was further developed to deal with random effect models in~\cite{van2016exploring}, or to evaluate the performance of dynamic prediction models in~\cite{schoop2011measures} where the information available from a longitudinal covariate is updated at each time point. Note also that other types of predictive accuracy measures exist, called discrimination measures, such as the C-index (see~\cite{harrell1996multivariable}, ~\cite{gerds2013estimating}) or the time dependent ROC curve and area under the curve (see for instance~\cite{heagerty2005survival}).

In this paper, the aim is to derive a predictive accuracy measure for recurrent events models where the focus is on predictiveness rather than discrimination. The quantity of interest is solely the expected cumulative mean number of recurrent events. Since no mean squared error measure, such as the Brier score, exists in the context of recurrent events, the goal of this work is to fill in this gap by deriving a new score of this type for recurrent events, which also accommodates for the presence of a terminal event. In this work, we show that this score reduces to the Brier score when only one event per individuals can occur and hence can be seen as a direct generalisation of the standard Brier score. Also, since our prediction criterion focuses on the marginal quantity of the expected cumulative number of recurrent events, it provides a summary score that takes into account the prediction of all recurrent events. In the context of a terminal event, it also incorporates the quality of prediction of the terminal event.

In Section~\ref{sec:predic_general}, we introduce the general prediction criterion for recurrent events, denoted $\widehat{\mathrm{MSE}}$.  In Sections~\ref{sec:RC} and~\ref{sec:terminal}, we separate the modelling assumptions in cases where no terminal events are observed and in the presence of a terminal event. In Section~\ref{sec:excum}, we present some existing estimators for the expected cumulative number of recurrent events. In Section~\ref{sec:theo}, we derive the main theoretical results of this paper. We first introduce a theoretical criterion and show that it can be decomposed into an inseparability term and an imprecision term, similarly to the results in~\cite{gerds2006consistent}. The former does not depend on the model and cannot be removed while the latter is exactly the mean squared error between the recurrent event process and the prediction model of the expected cumulative mean number of recurrent events. We then show that our prediction criterion asymptotically converges towards the theoretical criterion. 
In Section~\ref{sec:link_Brier}, we demonstrate that when individuals can only experience one event, our prediction criterion is equivalent to the standard Brier score. In Section~\ref{sec:simu1}, a simulation study is conducted. First, the decomposition between inseparability and imprecision terms is illustrated. As the inseparability is, by far, the dominant term, we then recommend to consider as a prediction score, the difference of $\widehat{\mathrm{MSE}}$s between the considered model and a reference model. Second, we illustrate how this prediction score can be used in order to compare prediction models. In Section~\ref{sec:AFdata}, the atrial fibrillation dataset is studied. We show that the model with dependence on prior counts (which is a multi-state model), stratified with respect to atrial fibrillation type, provides the best prediction performance among all other models considered.
 
%
%
%
%
%
%
%

\section{Prediction criterion for the expected cumulative number of recurrent events}\label{sec:predic}

\subsection{The prediction criterion in the general framework}\label{sec:predic_general}

In this section we present a prediction criterion for two different recurrent event settings under right-censoring. A scenario with right-censoring only (Section~\ref{sec:RC}) and a scenario with the inclusion of a terminal event (Section~\ref{sec:terminal}) are investigated. In each case, a counting process of interest $N^*(t)$ is defined which counts the number of recurrent events that have occurred before time $t$. We assume that a multivariate external time dependent covariate vector $X(t)$ (see~\cite{kalbfleisch_book} for the definition of external covariates) is observed and we define $\mathcal M$ a class of bounded functions depending on $t$ and $X(t)$. For each $t\geq 0$, we define $\mathcal X_t$, the support of the process $\{X(u) : 0\leq u\}$ at the time point $u=t$. In both scenarios, censoring can occur such that the observed recurrent event process is $N(t):=N^*(t\wedge C)$, where $C$ is a censoring variable and $a \wedge b$ represents the minimum between $a$ and $b$. We also define $\tau$ the endpoint of the study. On the basis of i.i.d. replications $(N_i(t),X_i(t) : 0\leq t\leq \tau)$, we will assume that an estimator $\widehat{\mu}\in\mathcal M$ of the expected cumulative number of recurrent events $\mu^*(t\mid X(t)):=\mathbb E[N^*(t)\mid X(u) : 0\leq u\leq t]$ is available. We will say that this estimator is consistent if there exists $\mu\in\mathcal M$ such that for all $t\leq \tau$,
\begin{align*}
\sup_{x\in\mathcal X_t} |\widehat\mu(t\mid x)-\mu(t\mid x)|\rightarrow 0, \text{ in probability as } n\to\infty.
\end{align*}
The main goal of this paper is to develop a new mean squared error criterion designed to evaluate the performance of this estimator in different scenarios.


In Sections~\ref{sec:RC} and~\ref{sec:terminal} we propose to evaluate the prediction ability of a given estimator $\widehat{\mu}\in\mathcal M$, through the following criterion: 
\begin{align}\label{eq:MSE}
\widehat{\mathrm{MSE}}(t,\widehat{\mu})&= \frac 1n \sum_{i=1}^n \left(\int_0^t\frac{dN_i(u)}{1-\hat{G}(u-\mid X_i(u))}-\widehat{\mu}(t\mid X_i(t)) \right)^2,
\end{align}
where $\hat G$ is an estimator of $G$, the conditional cumulative distribution function of the censoring variable $C$ given $X(\cdot)$. The notation $\hat{G}(u-\mid X_i(u))$ indicates the left limit of the function $\hat{G}$ at $u$. We will assume uniform consistency of this censoring estimator in the following way.
\begin{assum}\label{as:unifc}
Let $\mathcal G$ be a model for the conditional censoring distribution. We say that $\hat{G}$ is a uniformly consistent estimator for $G\in\mathcal G$ if for all $t\leq \tau$,
\[\sup_{x\in\mathcal X_t} |\hat G(t\mid x)-G(t\mid x)|\rightarrow 0, \text{ in probability as } n \rightarrow \infty.\]
\end{assum}
Presentations of different estimators for $G$, depending on the considered scenario, are discussed in Sections~\ref{sec:RC} and~\ref{sec:terminal}.

We now introduce a theoretical criterion that would be available if the censoring distribution was known. For some function $\mu\in\mathcal M$, let:
\begin{align}\label{eq:MSEtheo}
\mathrm{MSE}(t,\mu) &= \mathbb E\left[\left(\int_0^t\frac{dN(u)}{1-G(u-\mid X(u))}-\mu(t\mid X(t))\right)^2 \right].
\end{align}
The crucial idea behind our criterion~\eqref{eq:MSE} comes from the fact that 
\begin{align}\label{eq:crucial}
\mathbb E\left[\int_0^t \frac{dN(u)}{(1-G(u-\mid X(u)))}\right]=\mathbb E[\mu^*(t\mid X(t))],
\end{align}
a relationship that is proved for each scenario in Sections~\ref{sec:RC} and~\ref{sec:terminal}. In Section~\ref{sec:theo}, we provide theoretical results for both scenarios that justify the appropriateness of the proposed criterion. In particular, Proposition~\ref{th:0} of Section~\ref{sec:theo} shows that the theoretical criterion can be decomposed in the following way:
\begin{align*}
\mathrm{MSE}(t,\mu) &=\mathbb E\left[\Big(\mu^*(t\mid X(t))-\mu(t\mid X(t))\Big)^2 \right]+A(t),
\end{align*}
with $A(t)$ not depending on $\mu$. The first term is an imprecision term and the second term is an inseparability (or residual) term that does not depend on the chosen model. It should be noted that this kind of result is similar to the imprecision/inseparability decomposition of the Brier score (see~\cite{gerds2006consistent}). On the other hand, Proposition~\ref{th:1} of Section~\ref{sec:theo} states that if $\widehat\mu$ is a consistent estimator for some $\mu\in\mathcal M$ then as $n$ tends to infinity, our empirical criterion~\eqref{eq:MSE} is asymptotically equivalent to the theoretical criterion~\eqref{eq:MSEtheo} evaluated at $\mu$. Then, this latter criterion will be minimal if $\mu=\mu^*$ from Proposition~\ref{th:0}. In case $\mu$ is not equal to $\mu^*$, a bias will occur which will be equal to the squared expectation between $\mu$ and $\mu^*$. The precise assumptions under which Proposition~\ref{th:0} and Proposition~\ref{th:1} hold are specified in each situation.  


%
In what follows, the two scenarios, with right censoring only and with this inclusion of a terminal event, are presented. It should be noted that the estimated quantity $\mu^*$, is based on the rate function, as defined in~\cite{lin2000semiparametric}  or~\cite{cook2007statistical}. In this modelling approach, the conditioning is performed on the covariates but, contrary to the intensity function, we do not condition on the entire history of the process. As a result, our criterion~\eqref{eq:MSE} works for general recurrent event processes and, in particular, they are not restricted to Poisson processes.


\subsection{Situations with right-censoring and no terminal event}\label{sec:RC}

As before, we introduce the counting process of interest $N^*(t)$ which counts the number of recurrent events that have occurred before time $t$. Let $X(t)$ be some multivariate external covariate (see~\cite{kalbfleisch_book}) that is allowed to depend on $t$. For each $t\geq 0$, we define $\mathcal X_t$, the support of the process $\{X(u) : 0\leq u\}$ at the time point $u=t$. The regression modelling approach considered here is defined by:
\begin{align}\label{eq:model0}
\mathbb E[dN^*(t)\mid X(t)]&=\lambda^*(t\mid X(t))dt,
\end{align} 
where $\lambda^*(t\mid X(t))$ is the true rate function. Note that this definition is very general and does not make any assumption on the recurrent event process such as independent increments or a Poisson assumption. See~\cite{lin2000semiparametric} for more details on this model.

Let $\int_0^t\lambda^*(u\mid X(u))du$ be the true cumulative rate function. In the absence of a terminal event, this cumulative function has a direct interpretation as the expected cumulative number of recurrent events given the covariate process: $ \int_0^t\lambda^*(u\mid X(u))du=\mu^*(t\mid X(t))$ where as defined at the beginning of Section~\ref{sec:predic}, $\mu^*(t\mid X(t))=\mathbb E[N^*(t)\mid X(u) : 0\leq u\leq t]$. In the presence of censoring, a variable $C$ is observed such that the observed recurrent event process is now $N(t)=N^*(t\wedge C)$. We assume independent censoring (see~\cite{lin2000semiparametric}) in the following way:
\begin{align*}
\mathbb E[dN^*(t)\mid X(t)]&=\mathbb E[dN^*(t)\mid C, X(t)].
\end{align*}
This assumption implies that $C$ does not convey any additional information on the probability of a jump of the recurrent event process. Under this assumption, we have
\begin{align}\label{eq:model0bis}
\mathbb E[dN(t)\mid I(C\geq t), X(t)]&=I(C\geq t) \lambda^*(t\mid X(t))dt,
\end{align}
where $I(\cdot)$ denotes the indicator function. This last equation justifies the use of our criterion~\eqref{eq:MSE} since $\mathbb E[dN(t)\mid X(t)]=(1-G(t-\mid X(t))) \lambda^*(t\mid X(t))dt$ and therefore Equation~\eqref{eq:crucial} holds.
Next, we assume Assumption~\ref{as:unifc} and we make the following additional assumption. 
\begin{assum}\label{as:1}
We assume that there exists a constant $\tau>0$ and a constant $c>0$ such that 
\begin{enumerate}
\item $\forall t\in[0,\tau]$, $\mathbb P[C\geq t\mid X(t)]\geq c$ almost surely,
\item $N(\tau)$ is almost surely bounded by a constant.
\end{enumerate}
\end{assum}
Note that condition~{\it 1.} was also assumed in~\cite{gerds2006consistent}. It is stronger than simply assuming $\mathbb P[C\geq \tau]\geq c$. Condition~{\it 2.}  is 
standard for recurrent event data, see for instance~\cite{lin2000semiparametric}. Finally, note that through Equation~\eqref{eq:model0bis}, conditions~{\it 1.} and~{\it 2.} imply that $\mathbb E[\mu^*(\tau\mid X(\tau))]<\infty$.

On the basis of i.i.d. replications $(N_i(t),X_i(t) : 0\leq t\leq \tau)$, let $\widehat{\mu}\in\mathcal M$ be an estimator of $\mu^*$ where $\mathcal M$ is a class of models that are assumed to be bounded. 
%
We propose to evaluate the prediction ability of this estimator through criterion $\widehat{\mathrm{MSE}}(t,\widehat{\mu})$ defined in Equation~\eqref{eq:MSE}. This criterion involves an estimator of $G$, the conditional cumulative distribution function of the censoring variable. If $C$ and $X(\cdot)$ are independent, one can estimate $G$ using the empirical cumulative distribution function of the censored variables since all these variables are observed. If $C$ depends on $X$ the conditional distribution of $C$ must also be modelled. This can be done using kernel based estimators such as the Nadaraya-Watson estimator for the binary response variable $I(C\leq t)$ or extensions of this model. For instance, in~\cite{hall1999methods}, a local logistic method and an adjusted Nadaraya-Watson estimator are proposed. If the dimension of the covariates that are assumed to depend on the censoring distribution is too large, then a dimension reduction technique can first be employed, for example through a Single-Index-Modelling approach (see~\cite{delecroix2003efficient}).

Theoretical results on the validity of this criterion are derived in Section~\ref{sec:theo}.
\subsection{Situations with right-censoring and a terminal event}\label{sec:terminal}

In this section, 
we introduce a terminal event $T^*$ such that the counting process of interest $N^*$ verifies (see e.g.~\cite{cook2007statistical}):
\begin{align}\label{eq:model1}
\mathbb E[dN^*(t)\mid I(T^*\geq t),X(t)]&=I(T^*\geq t)\lambda^*(t\mid X(t))dt,
\end{align} 
where $\lambda^*(t\mid X(t))$ is the true rate function. The difference in Model~\eqref{eq:model1} with respect to Model~\eqref{eq:model0} is that the recurrent event process $N^*$ is stopped by the terminal event. This often occurs  in real-data analysis where death is typically the terminal event. Under this model, we observe that
\begin{align*}
\mu^*(t\mid X(t))=\mathbb E[N^*(t)\mid X(u) : 0\leq u\leq t]=\int_0^t S(u \mid X(u))\lambda^*(u\mid X(u))du,
\end{align*}
where $S(t \mid X(t)):=\mathbb P[T^*\geq t\mid X(t)]$ is the conditional survival function of the terminal event. This implies that in order to define an estimator of $\mu^*(t\mid X(t))=\mathbb E[N^*(t)\mid X(u) : 0\leq u\leq t]$ one usually needs to also model the hazard rate for the terminal event and to derive an estimator of the conditional survival function. As a result our prediction criterion will both take into account the predictive performance of the survival function and of the rate function of $N^*$ since, if one of those two estimators behaves poorly, the resulting estimator for $\mu^*$ is likely to perform badly as well.

The independent censoring assumption is now expressed in the following way:
\begin{align*}
\mathbb E[dN^*(t)\mid T^*, X(t)]&=\mathbb E[dN^*(t)\mid T^*, C, X(t)].
\end{align*}
We denote $T=T^*\wedge C$ the minimum between terminal event and censoring, $Y(t)=I(T\geq t)$ the observed at-risk process and $N(t)=N^*(T\wedge t)$ the observed counting process. Under the independent censoring assumption, it can be shown that
\begin{align}\label{eq:model1bis}
\mathbb E[dN(t)\mid Y(t),X(t)]&=Y(t)\lambda^*(t\mid X(t))dt.
\end{align} 
We assume Assumption~\ref{as:unifc} and we make the following additional assumption. 
\begin{assum}\label{as:2}
We assume that there exists a constant $\tau>0$ and a constant $c>0$ such that 
\begin{enumerate}
\item $\forall t\in[0,\tau]$, $\mathbb P[T\geq t\mid X(t)]\geq c$ almost surely,
\item $N(\tau)$ is almost surely bounded by a constant.
\end{enumerate}
We also assume that $T^*$ is independent of $C$ conditionally on $X(\cdot)$.
\end{assum}
Those conditions are standard in the context of regression for recurrent events with a terminal event, see~\cite{ghosh2002marginal} for example. Using Equality~\eqref{eq:model1bis} one can easily observe that $\mathbb E[dN(t)\mid X(t)]=S(t\mid X(t))(1-G(t-\mid X(t))\lambda^*(t\mid X(t))dt$ under the independent censoring hypothesis in Assumption~\ref{as:2}. We then directly see that Equation~\eqref{eq:crucial} holds.

On the basis of i.i.d. replications $(N_i(t),X_i(t) : 0\leq t\leq \tau)$, let $\widehat{\mu}\in\mathcal M$ be an estimator of $\mu^*$ where $\mathcal M$ is a class of models that are assumed to be bounded. 
We propose to evaluate the prediction ability of this estimator through criterion $\widehat{\mathrm{MSE}}(t,\widehat{\mu})$ defined in Equation~\eqref{eq:MSE}. This criterion involves an estimator of $G$, the conditional cumulative distribution function of the censoring variable. If $C$ and $X(\cdot)$ are independent, one can estimate $G$ using the Kaplan-Meier estimator by considering $C$ to be the variable of interest that is incompletely observed due to the terminal event $T^*$. If $C$ depends on $X$ the conditional distribution of $C$ must also be modelled. Several possible models are presented in~\cite{gerds2006consistent} such as the Cox model, the Aalen additive model, or the kernel type model of~\cite{dabrowska1989uniform}. Alternatively, a single-index approach for right-censored data, as in~\cite{bouaziz2010conditional}, or the random survival forest method developed in~\cite{ishwaran2008random} can be used.

Theoretical results on the validity of this criterion are derived in Section~\ref{sec:theo}.

\subsection{Examples of estimators for the expected cumulative number of recurrent events}\label{sec:excum}

In this section we present some estimators for the expected cumulative mean number. In order to achieve this goal, one possibility is to first model the rate function $\lambda^*$ and then to use a plug-in estimator to derive the final estimator of $\mu^*$. For example, the Cox (see~\cite{cox1972regression}), Aalen (see~\cite{scheike2002additive}) or Accelerated Failure Time (see~\cite{lin1998accelerated}) models  can be used for the estimation of $\lambda^*$.  
We note $\widehat\lambda$ and $\widehat\Lambda$ such estimators of $\lambda$ and $\Lambda$. In the context of right-censored data and no terminal event, the estimator of $\mu^*$ can then be expressed in the following way (see~\cite{cook2007statistical}):
\begin{align}\label{eq:ex_noterm}
\widehat\mu(t\mid X(t))=\int_0^t d\widehat\Lambda(u\mid X(u)).
\end{align}

In the presence of a terminal event, a common approach is to first model the hazard rate of the terminal event (using again a Cox model for instance) and to derive an estimator of the survival function $\widehat S(t\mid X(t))=\exp(-\int_0^t \widehat{\lambda}^{T^*}(u\mid X(u))du)$ where $\widehat{\lambda}^{T^*}$ is the estimator of the hazard rate of the terminal event. Then, the final estimator of $\mu^*$ is (see~\cite{cook2007statistical} and~\cite{andersen2019modeling}):
\begin{align}\label{eq:ex_term}
\widehat\mu(t\mid X(t))=\int_0^t \widehat S(u\mid X(u))d\widehat\Lambda(u\mid X(u)).
\end{align}

Alternative approaches that directly model $\mu^*$ also exist. In~\cite{ghosh2002marginal} the authors consider the following Cox type model: $\mu^*(t\mid X(t))=\mu_0(t)\exp(X(t)'\beta)$. A more general approach consists in using a Single-Index-Model for estimating $\mu^*$: in~\cite{bouaziz2015semiparametric} the authors assume the existence of a nonparametric function $g$ and a parameter $\beta$ such that $\mu^*(t\mid X(t))=g(X(t)'\beta)$. 
Those two approaches provide a direct estimator of the quantity of interest $\mu^*$. However, in the presence of a terminal event, it is no longer possible to disentangle the effects on the recurrent event process or the terminal event. As a result, the regression parameters should be interpreted with caution. See also~\cite{cook2007statistical} for a discussion about this issue.

\section{Theoretical results}\label{sec:theo}


In this section we provide theoretical results on the proposed criterion~\eqref{eq:MSE} in the two different contexts that have been previously considered: the scenario with right-censoring only and the scenario with right-censoring and a terminal event. Two results are obtained. The first one is concerned with the theoretical criterion~\eqref{eq:MSEtheo}. It shows that this criterion applied to a function $\mu\in\mathcal M$ reduces to the mean squared error between $\mu$ and $\mu^*$ and a term that does not depend on $\mu$. The second result shows the asymptotic consistency between $\widehat{\mathrm{MSE}}(t,\hat\mu)$ and $\mathrm{MSE}(t,\mu)$ when the estimator $\widehat{\mu}$ is an asymptotically consistent estimator of $\mu$. 

 \begin{proposition}\label{th:0}
For the scenarios with right-censoring only (Section~\ref{sec:RC}) and with right-censoring and terminal event (Section~\ref{sec:terminal}) we respectively assume Assumption~\ref{as:1}, Assumption~\ref{as:2}. In both scenarios we also assume independent censoring. We then have for $\mu\in\mathcal M$,
\begin{align}\label{eq:th0}
\mathrm{MSE}(t,\mu) &=\mathbb E\left[\Big(\mu^*(t\mid X(t))-\mu(t\mid X(t))\Big)^2 \right]+A(t),
\end{align}
where $A(t)\geq 0$ for all $t\geq 0$ and $A(t)$ does not depend on $\mu$.
 \end{proposition}

\begin{proposition}\label{th:1}
For the scenarios with right-censoring only (Section~\ref{sec:RC}) and with right-censoring and terminal event (Section~\ref{sec:terminal}) we respectively assume Assumption~\ref{as:1} or Assumption~\ref{as:2}. In both scenarios we also assume Assumption~\ref{as:unifc} and independent censoring. 
Then, if the estimator $\widehat{\mu}\in\mathcal M$ is consistent for $\mu\in\mathcal M$, we have
\begin{align*}
\sup_{t\leq \tau} \left|\widehat{\mathrm{MSE}}(t,\hat\mu)-\mathrm{MSE}(t,\mu)\right|\to 0,\quad \text{in probability as } n\to \infty.
\end{align*}
\end{proposition}
The proofs are provided in the Appendix, in Sections~\ref{proof:th0} and~\ref{proof:th1}, with an explicit expression of $A(t)$ in both scenarios.

\section{Link with the Brier score}\label{sec:link_Brier}

The Brier score (see~\cite{gerds2006consistent}) is a popular criterion to evaluate the prediction performance of a regression model for the conditional survival function in the context of right-censoring when only a single event can be observed per individual. We show in this section that if we use our criterion when individuals can only experience one event at most, then our theoretical criterion denoted $\mathrm{MSE}'$ and defined as follows
\begin{align}\label{eq:MSEtheo_Brier}
\mathrm{MSE}'(t,\pi) &= \mathbb E\left[\left(1-\int_0^t\frac{dN(u)}{1-G(u-\mid X(u))}-\pi(t\mid X(t))\right)^2 \right],
\end{align} 
where $\pi=1-\mu$, reduces to the theoretical Brier score up to a term that does not depend on the model $\mathcal M$. Note that when individuals can only experience one event, the recurrent event process reduces to $N^*(t)=I(T^*\leq t)$ and $\mu^*(t\mid X)=\mathbb E[N^*(t)\mid X]$ is the conditional cumulative distribution function of $T^*$. Since the Brier score has been designed for the prediction of the conditional survival function, we have simply rewritten our criterion in Equation~\eqref{eq:MSEtheo} such that $\pi$ represents the model for the conditional survival function.

We first recall that the theoretical Brier score is defined as (see Equation~(1) from~\cite{gerds2006consistent}):
\begin{align*}
\mathrm{MSE^{Brier}}(t,\pi)= \mathbb E\,[(I(T^*>s)-\pi(t\mid X))^2],
\end{align*}
where the expectation is taken with respect to the joint distribution of $T^*$ and $X$. For simplicity, the covariate $X$ is not time dependent in the formula, as presented in the paper of~\cite{gerds2006consistent}, but the results presented in this section are still valid for time dependent covariates. In their work, the authors show similar results as Propositions~\ref{th:0} and~\ref{th:1} of the present paper when the aim is to provide a prediction of the survival function $S(t\mid X)=\mathbb P[T^*>t\mid X]$. Note that we have suppressed the dependency with respect to $S$ in the definition of the Brier score to stay consistent with the notations used throughout this paper. Also, in the definition of the Brier score, $\pi$ plays the role of $1-\mu$ in the present paper, that is, it is the limiting function of a proposed conditional survival estimator $\hat S(t\mid X)$. We have the following result.


\begin{proposition}\label{th:2}
We assume that only one event per individual can be experienced, that is, $N(t)=I(T\leq t,\Delta=1)$, with $T=T^*\wedge C$ is the observed time, $T^*$ is the true event time, $C$ is the censoring variable and $\Delta=I(T^*\leq C)$ is the censoring indicator. Then, under independent censoring, we have:
\begin{align*}
\mathrm{MSE}'(t,\pi) 
 & =  \mathrm{MSE^{Brier}}(t,\pi)  + B(t),
\end{align*}
where $B(t)\geq 0$ for all $t\geq 0$ and $B(t)$ does not depend on $\pi$.
\end{proposition}

The proof is provided in the Appendix, in Section~\ref{proof:th3}, with an explicit expression of $B(t)$. Since $B(t)$ does not depend on the model $\pi$, those two criterions are completely equivalent. In particular, if one considers the difference of a regression model to a reference model, such as a model that does not include covariates, then the two criterions will provide exactly the same values since the $B(t)$ term will cancel out in the difference. As we will see in the next section, comparing a model to a reference is typically what we recommend in practice. Since recurrent events are a generalisation of the single-event per individual situation, our criterion can be seen as an extension of the Brier score for recurrent events.

\section{Simulations}

\subsection{A scenario with right-censoring and no terminal events}\label{sec:simu1}

For $i=1,\ldots,n$, we first simulate a two-dimensional covariate vector $X_i=(X_{i,1},X_{i,2})^{\top}$ with $X_{i,1}$ a Bernoulli variable with parameter $0.5$ and $X_{i,2}$ a Gaussian variable with expectation $2$ and standard deviation $0.5$. Conditional on $X_i$, 
the recurrent events are generated from a non-homogeneous Poisson process with rate $\lambda^*(\cdot\mid X_i)$ that follows a Cox model with Weibull baseline and a two-dimension time independent covariate. More specifically, 
\begin{align*}
\lambda^*(t\mid X_i)=\lambda_0(t)\exp(\theta_0^{\top}X_i), \quad\lambda_0(t) = \frac{\alpha}{\beta} \left(\frac{t}{\beta}\right)^{\alpha-1},
\end{align*}
with $\alpha=2$ the shape parameter, $\beta=0.39$ the scale parameter and $\theta_0=(\log (2),\log (0.5))^{\top}$. Recurrent events generated under this simulation setting follow Equation~\eqref{eq:model0} with the true expected number of recurrent events equal to
\begin{align}\label{eq:simu_mu}
\mu^*(t\mid X_i) = \int_0^t \lambda^*(t\mid X_i) du=\left(\frac{t}{\beta}\right)^{\alpha}\exp(\theta_0^{\top}X_i).
\end{align}
 We further simulate a censoring variable $C_i$ that follows a uniform distribution on $[0,3]$. Using those parameters, we observe $0$ or $1$ recurrent event for $30\%$ of the individuals, less or equal than $5$ events for $54\%$ of the individuals, and less or equal than $12$ events for $77\%$ of the individuals. On average, we observe approximately $8$ recurrent events per individual.

\begin{figure}[!htb]
\centering
\resizebox{\textwidth}{!}{ 
\begin{tabular}{c}
\includegraphics[scale=0.1]{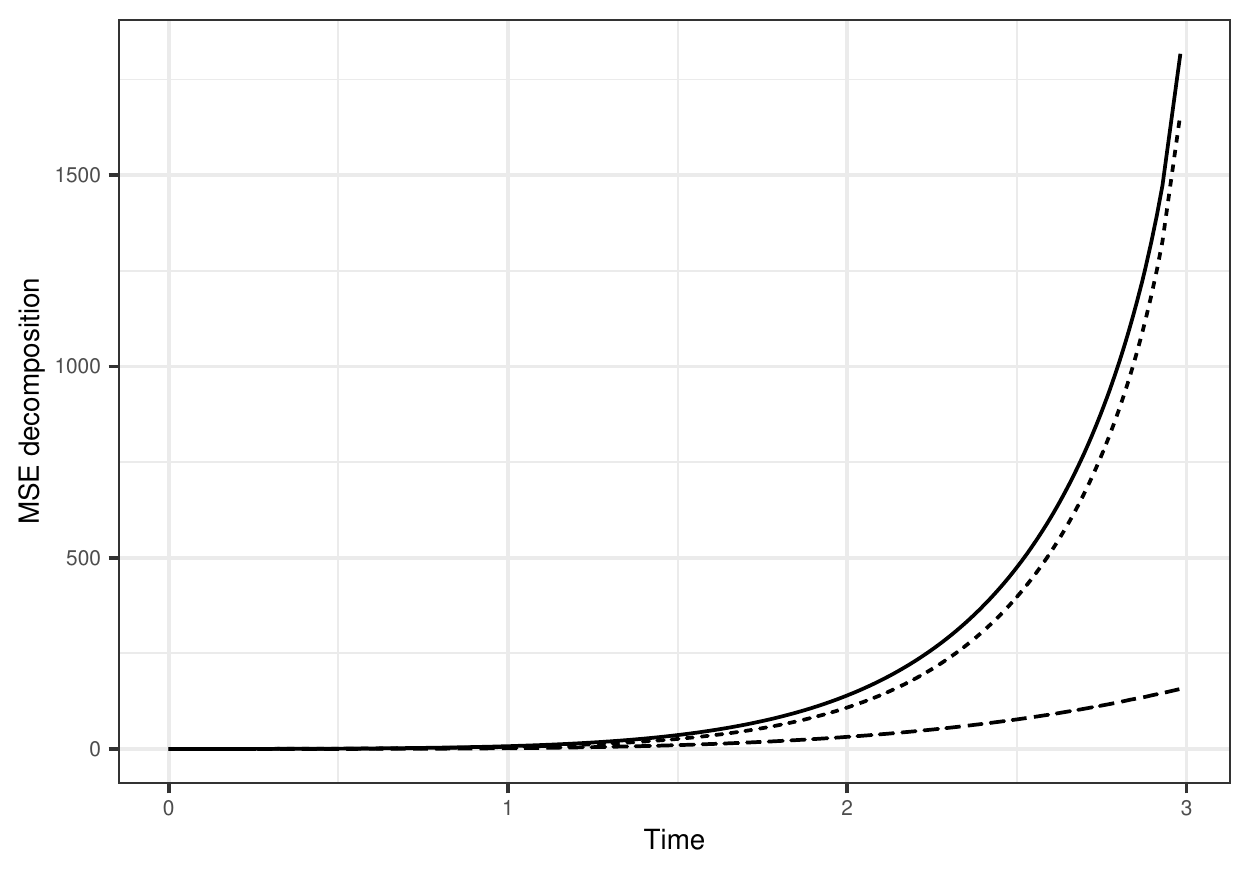}
\end{tabular}}
\caption{\footnotesize Decomposition of the MSE (solid line) in Proposition~\ref{th:0} as the sum between the the inseparability term $A(t)$ (dotted line) and the imprecision term (dashed line). The data were simulated from a Cox model with two covariates and the expected cumulative number of recurrent events was predicted using the Nelson-Aalen estimator. The train sample ($n_{\text{train}}=200$) is used for the computation of the Nelson-Aalen estimator, the test sample ($n_{\text{test}}=1,000$) is used for the computation of the MSE.}
\label{fig:plot1}
\end{figure}

Based on a single simulated sample, we first illustrate Propositions~\ref{th:0} and~\ref{th:1} when the class of models $\mathcal M$ that contains $\mu$ assumes no effect of the covariates on the occurrence of the recurrent events. 
For this purpose, we independently simulate a training and a test samples. The training sample is used to compute $\hat \mu^{\text{train}}$ based on Equation~\eqref{eq:ex_noterm} where $\hat\Lambda$ does not depend on $X$ and is simply the Nelson-Aalen estimator: 
\begin{align*}
\hat\Lambda(t)=\frac 1n \sum_{i=1}^n \int_0^t \frac{dN_i(u)}{1-\hat G(u-)},
\end{align*}
with $\hat G$ the Kaplan-Meier estimator of $C$. \sloppy We then compute $\widehat{\mathrm{MSE}}(t,\widehat{\mu}^{\text{train}})$ from Equation~\eqref{eq:MSE} based on the test sample of size $n_{\text{test}}$, that is, the computation is performed on a sample $(N_1(\cdot),X_1),\ldots,(N_{n_{\text{test}}}(\cdot),X_{n_{\text{test}}})$ independent of the training sample. This quantity should provide an accurate estimation of $\mathrm{MSE}(t,\mu)$ from Proposition~\ref{th:1}. We then compute the imprecision term $\mathbb E\left[\Big(\mu^*(t\mid X)-\mu(t\mid X)\Big)^2 \right]$ in Equation~\eqref{eq:th0} using the true value of $\mu^*(t\mid X(t))$, replacing $\mu$ by $\hat \mu^{\text{train}}$ and replacing the expectation by its empirical sum. In other words, we compute
\begin{align*}
\frac 1{n_{\text{test}}} \sum_{i=1}^{n_{\text{test}}} \Big(\mu^*(t\mid X_i)-\hat \mu^{\text{train}}(t\mid X_i)\Big)^2,
\end{align*}
which should give a very accurate estimation of the imprecision term. The $A(t)$ term is exactly computed based on its explicit expression (see Section~\ref{sec:A_comput} in the Appendix). The decomposition of the MSE between its imprecision and inseparability terms is displayed in Figure~\ref{fig:plot1} using $n_{\text{train}}=200$ and $n_{\text{test}}=1,000$. In Equation~\eqref{eq:MSE} the Kaplan-Meier estimator of $G$ was computed from the combination of the training and test samples. The solid line represents the estimated MSE while the dotted and dashed lines represent the inseparability and imprecision terms, respectively. The inseparability term is seen to be very close to the MSE. In contrast, the imprecision term, which clearly is not null here since the estimated model uses no covariates, is relatively small as compared to the other two terms. This plot suggests that it might be difficult to compare different models as the inseparability term is dominant in the decomposition of the MSE, which implies that two MSEs computed from two different models will tend to look very similar (for instance, for $t=2.5$, the inseparability term represents approximately $84\%$ of the value of the MSE). As a result, we advocate the use of a reference or null model and to compute the score of a given model as the difference between the MSE of the reference and the MSE of this model. Therefore, this score will represent the prediction gain of the model as compared to the null model. A typical choice of the null model is the one that uses no covariates. The score is defined as
\begin{align}\label{eq:score}
\text{Score}(t,\widehat{\mu},\widehat{\mu_0}) = \widehat{\mathrm{MSE}}(t,\widehat{\mu_0}) - \widehat{\mathrm{MSE}}(t,\widehat{\mu}),
\end{align}
where $\widehat{\mu}$ is the evaluated model and $\widehat{\mu_0}$ is the reference model. Those models will usually be implemented based on a training sample. The idea behind this score is that the inseparability term will cancel out in the difference, and the score is therefore equal to the difference between the imprecision terms of the two models.

\begin{figure}[!htb]
\centering
\resizebox{\textwidth}{!}{ 
\begin{tabular}{c}
\includegraphics[width=0.1\textwidth]{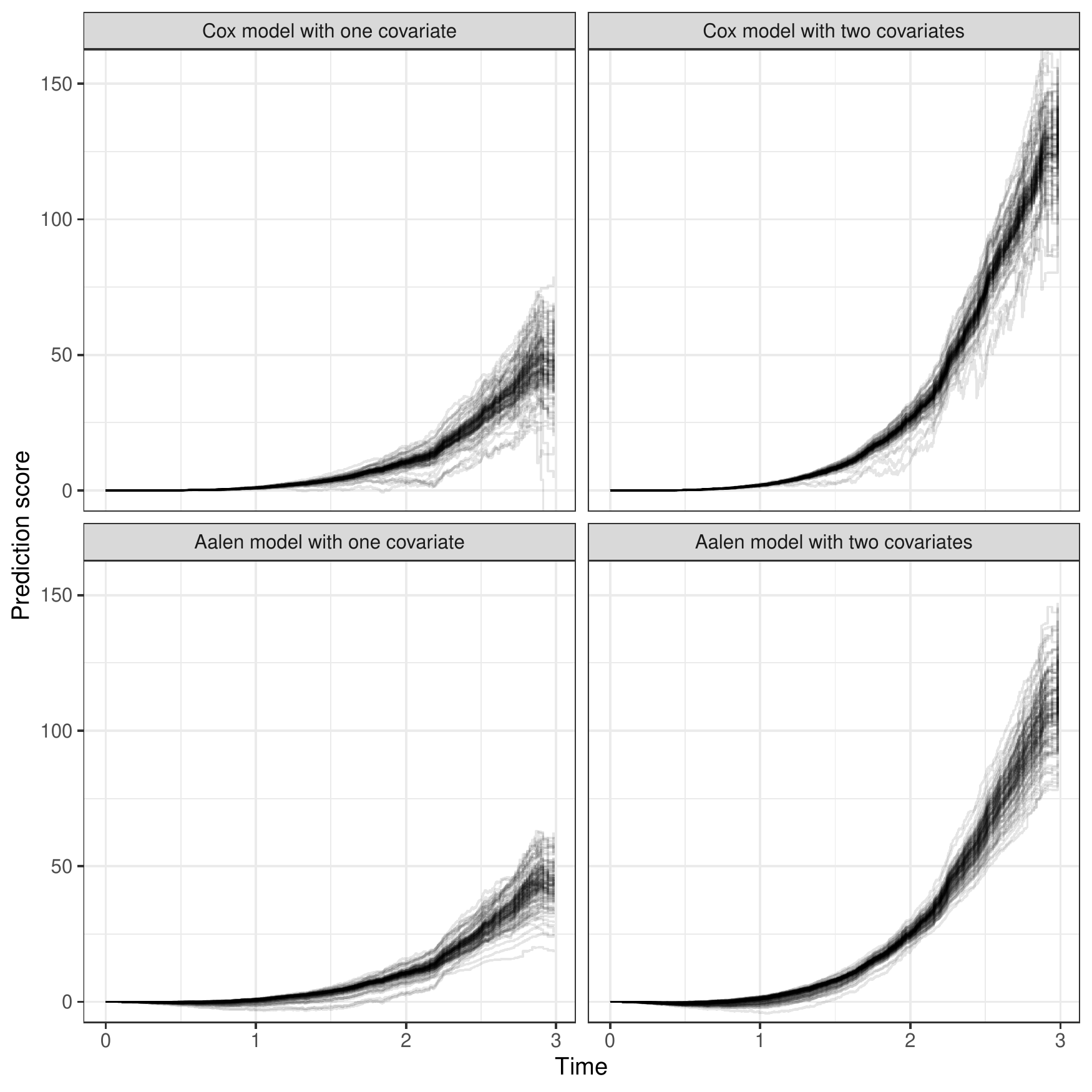}
\end{tabular}}
\caption{\footnotesize Prediction scores (see Equation~\eqref{eq:score}) using four different models. The data were generated from a Cox model with two covariates and the expected cumulative number recurrent of events was predicted using the Cox model with one covariate, the Cox model with two covariates, the Aalen model with one covariate, the Aalen model with two covariates, respectively. The reference model uses no covariates and was estimated from the Nelson-Aalen estimator. The prediction scores are computed for $100$ training samples of size $n_{\text{train}}=50$ and a unique test sample of size $n_{\text{test}}=1,000$.}
\label{fig:plot2}
\end{figure}

An illustration of this score is presented in Figure~\ref{fig:plot2}. Using the same simulation setting as before, we compare the performance of four different models based on the Cox and Aalen models, implemented using either only the first covariate $X_{i,1}$ or the two covariates $X_{i,1}$ and $X_{i,2}$. Figure~\ref{fig:plot2} displays the prediction scores for $100$ training samples of size $50$ and a unique test sample of size $1,000$. The reference model is the one that uses no covariates and is estimated from the Nelson-Aalen estimator. Roughly, we see that all models have a better prediction performance than the Nelson-Aalen estimator as time increases especially 
from time equal to $1.5$ and time equal to $2$, for the models with one covariate and the models with two covariates, respectively. The models with two covariates clearly outperform the ones with one covariate with a slightly better performance for the Cox model as compared to the Aalen model. This is further illustrated in Table~\ref{tab:1} where we compare the mean score of those four different models based on $500$ training samples of size $20$ and $50$ and one single test sample of size $1,000$. We clearly observe that the correctly specified Cox model with two covariates outperforms all other models on average, for all time points and sample sizes. However, it tends to have a slightly bigger standard deviation, especially for $n_{\text{train}}=20$ and $t=2$ or $t=2.9$.

\begin{table}[ht]
\centering
\resizebox{\textwidth}{!}{ 
\begin{tabular}{|l|ccc|ccc|}
  \hline
  & \multicolumn{3}{c|}{$n_{\text{train}}=20$} & \multicolumn{3}{c|}{$n_{\text{train}}=50$}\\
 & $t=1$ & $t=2$ & $t=2.9$ & $t=1$ & $t=2$ & $t=2.9$ \\
 \hline
Cox one cov. & $0.89$ $(0.33)$ & $9.17$ $(6.04)$ & $40.6$ $(54.93)$ & $0.96$ $(0.15)$ & $10.06$ $(2.19)$ & $46.93$ $(16.06)$ \\ 
  Cox two cov. & $1.8$ $(0.56)$ & $24.69$ $(8.58)$ & $119.11$ $(45.56)$ & $1.95$ $(0.17)$ & $26.33$ $(2.54)$ & $127.52$ $(17.73)$ \\ 
  Aalen one cov. & $0.14$ $(1.33)$ & $9.53$ $(4.4)$ & $42.72$ $(16.27)$ & $0.34$ $(0.82)$ & $10.09$ $(1.99)$ & $44.7$ $(9.21)$ \\ 
  Aalen two cov. & $0.44$ $(1.76)$ & $24.18$ $(4.8)$ & $103.78$ $(25.28)$ & $0.51$ $(1.22)$ & $25.13$ $(1.9)$ & $109.36$ $(14.21)$ \\ 
   \hline
\end{tabular}}
\caption{{\small Means and standard deviations (in bracket) 
over $500$ simulations for the prediction score of the expected number of recurrent events. Large values indicate better predictive performances.}} 
\label{tab:1}
\end{table}

%

\subsection{A scenario with right-censoring and a terminal event}

We now consider a simulation scenario which also includes a terminal event. The recurrent event process and its covariates are  simulated in the same manner as in the previous section, with the same parameter values. The censoring variable is simulated following a uniform variable on $[0,8]$. The terminal event is simulated according to a Cox model with baseline following a Weibull distribution with shape parameter equal to $5$ and scale parameter equal to $1.8$. This Cox model also includes the same two covariates as for the recurrent event process with the same effects on the hazard function (i.e. the effects are equal to $\log(2)$, $\log(0.5)$ for the Bernoulli and Gaussian covariates respectively). This setting leads to $8.5$ events per individual on average, with $26\%$, $50\%$, $77\%$ of individuals that experience less than or equal to $3$, $7$ and $12$ events, respectively.  On average $28\%$ of individuals are censored. 

We estimate the expected cumulative number of recurrent events based on Equation~\eqref{eq:ex_term} where in the formula, the Breslow estimator is used to estimate the conditional survival function of the terminal event, if the estimation model for the terminal event includes covariates. In other words:
\begin{align*}
\widehat S(t\mid X) = \exp\left(-\int_0^t \exp(X^{\top}\hat \beta^{T^*})d\widehat{\Lambda}_0^{T^*}\right),
\end{align*}
with $\hat\beta^{T^*}$ is the estimated regression parameter from the Cox model for the terminal event and $\hat\Lambda_{0}^{T^*}$ its corresponding baseline estimator known as the Breslow estimator. If the terminal event model does not contain any covariates, then the Kaplan-Meier estimator is used instead. 
As previously, we use the score defined in Equation~\eqref{eq:score} to evaluate the quality of prediction of a model where the reference model $\widehat\mu_0$ is defined as
\begin{align}\label{eq:ex_term_0}
\widehat\mu_0(t)=\int_0^t \hat S(u)d\widehat\Lambda(u),
\end{align}
with $\hat S$ the Kaplan-Meier estimator of the terminal event and $\widehat\Lambda$ the Nelson-Aalen estimator of the recurrent event process. The same score is also used for the prediction of the survival function with the Kaplan-Meier estimator as the reference model. We consider four different regression models: a correctly specified model that includes the two covariates for both the recurrent event process and the terminal event in two Cox models, a model where the Gaussian covariate is missing for the Cox model of the terminal event (but the Cox model of the recurrent event is correctly specified) and a model where the Gaussian covariate is missing for both Cox models. 

\begin{figure}[!htb]
\captionsetup[subfigure]{labelformat=empty}
\centering
\begin{subfigure}[b]{0.75\textwidth}
   \includegraphics[width=0.8\textwidth]{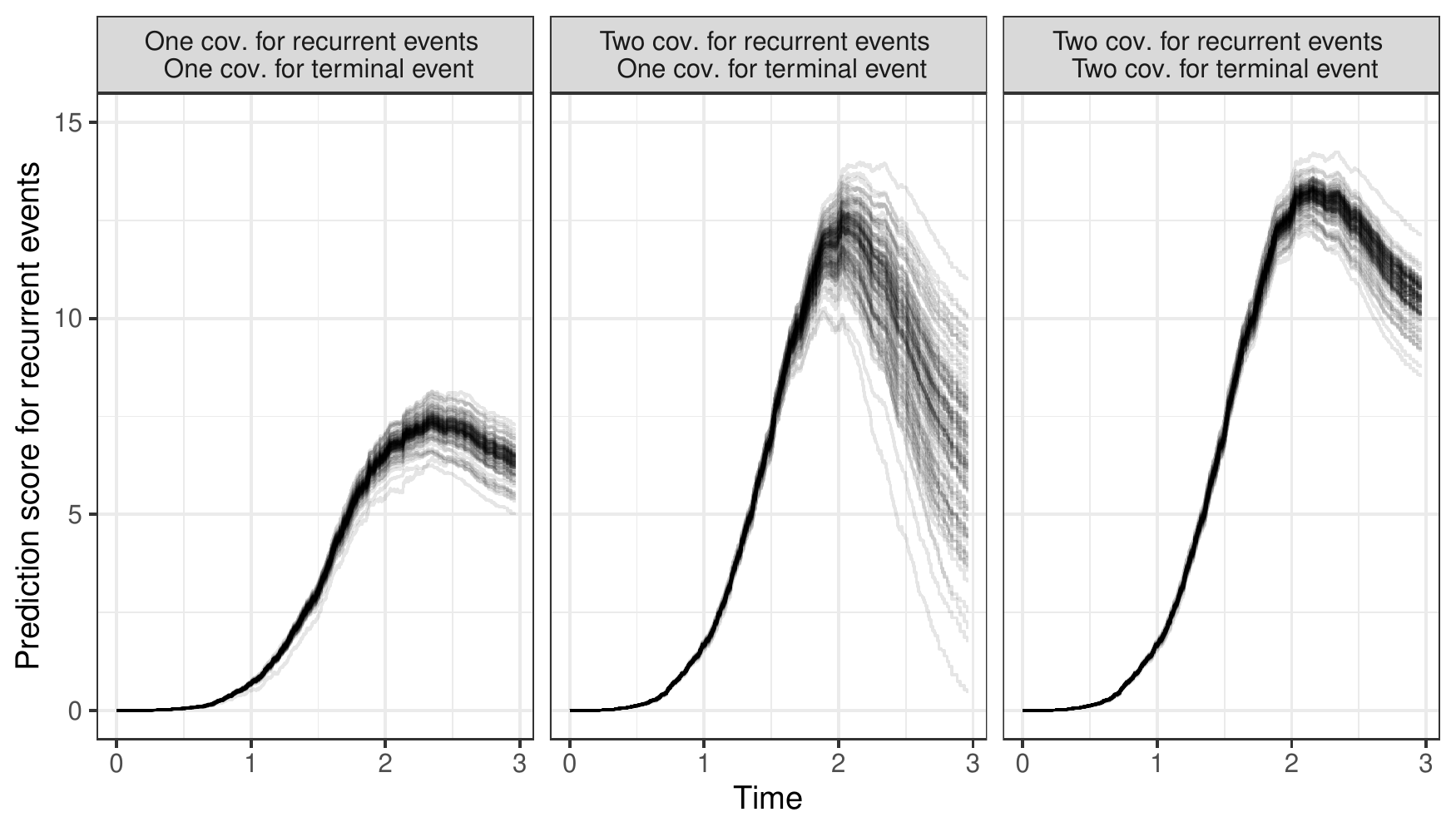}
   \caption{}
\end{subfigure}

\begin{subfigure}[b]{0.75\textwidth}
   \includegraphics[width=0.8\textwidth]{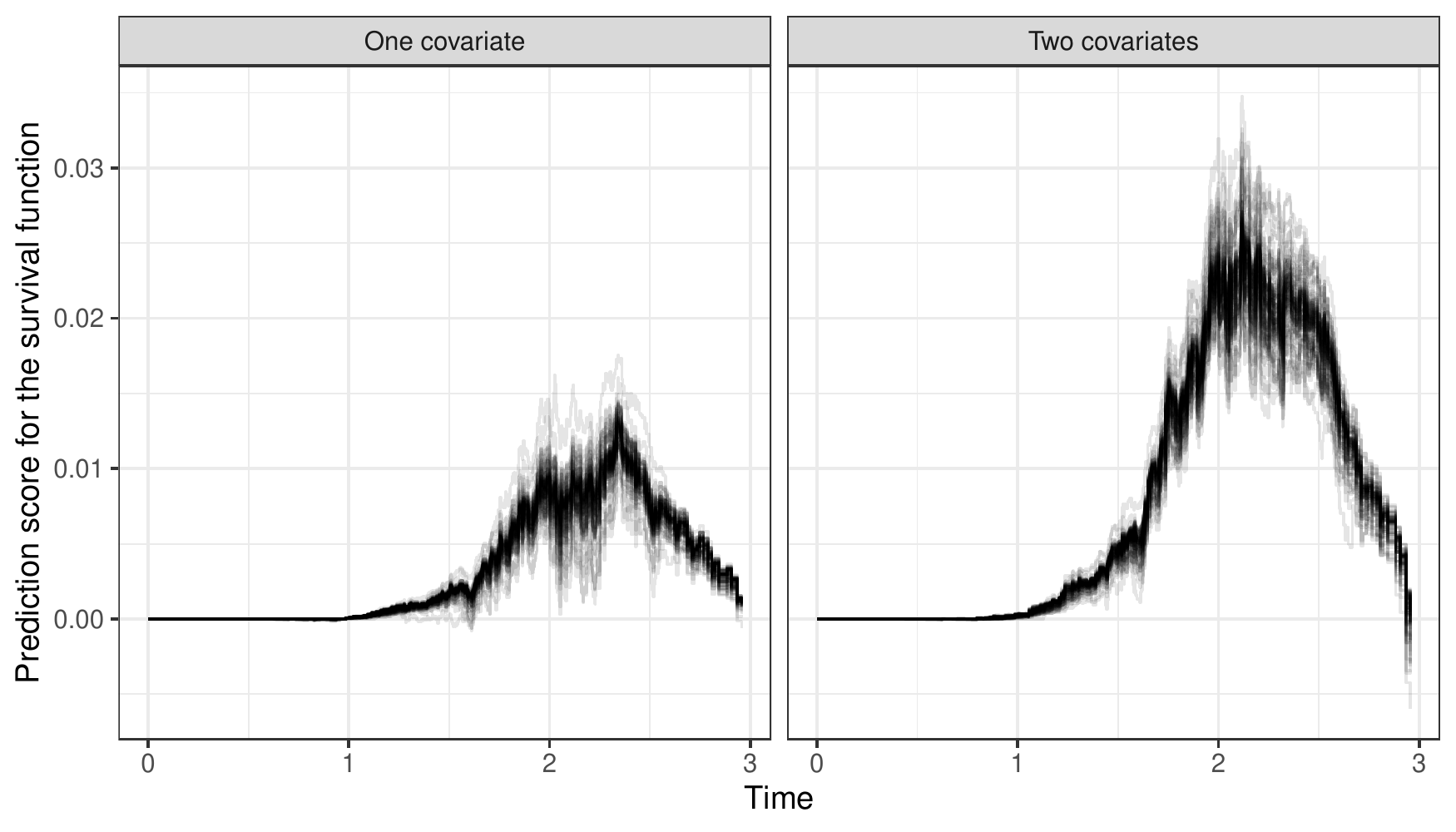}
   \caption{}
\end{subfigure}
\caption{{\footnotesize Prediction scores for the recurrent events (top panels) and the survival function (bottom panels) using different models. The data were generated from a Cox model with two covariates ($n=200$) for the recurrent event process and with the same two covariates for the terminal event. The expected cumulative number recurrent of events and the survival function of the terminal event were predicted using the Cox model with one or two covariates. The reference model uses no covariates and was estimated from the non-parametric estimator in Equation~\eqref{eq:ex_term_0} and from the Kaplan-Meier estimator in the top and bottom panels, respectively. The prediction scores are computed for $100$ training samples of size $n_{\text{train}}=800$ and a unique test sample of size $n_{\text{test}}=1,000$.}}
\label{fig:plot3}
\end{figure}

\begin{figure}[!htb]
\centering
\resizebox{\textwidth}{!}{ 
\begin{tabular}{c}
\includegraphics[width=0.06\textwidth]{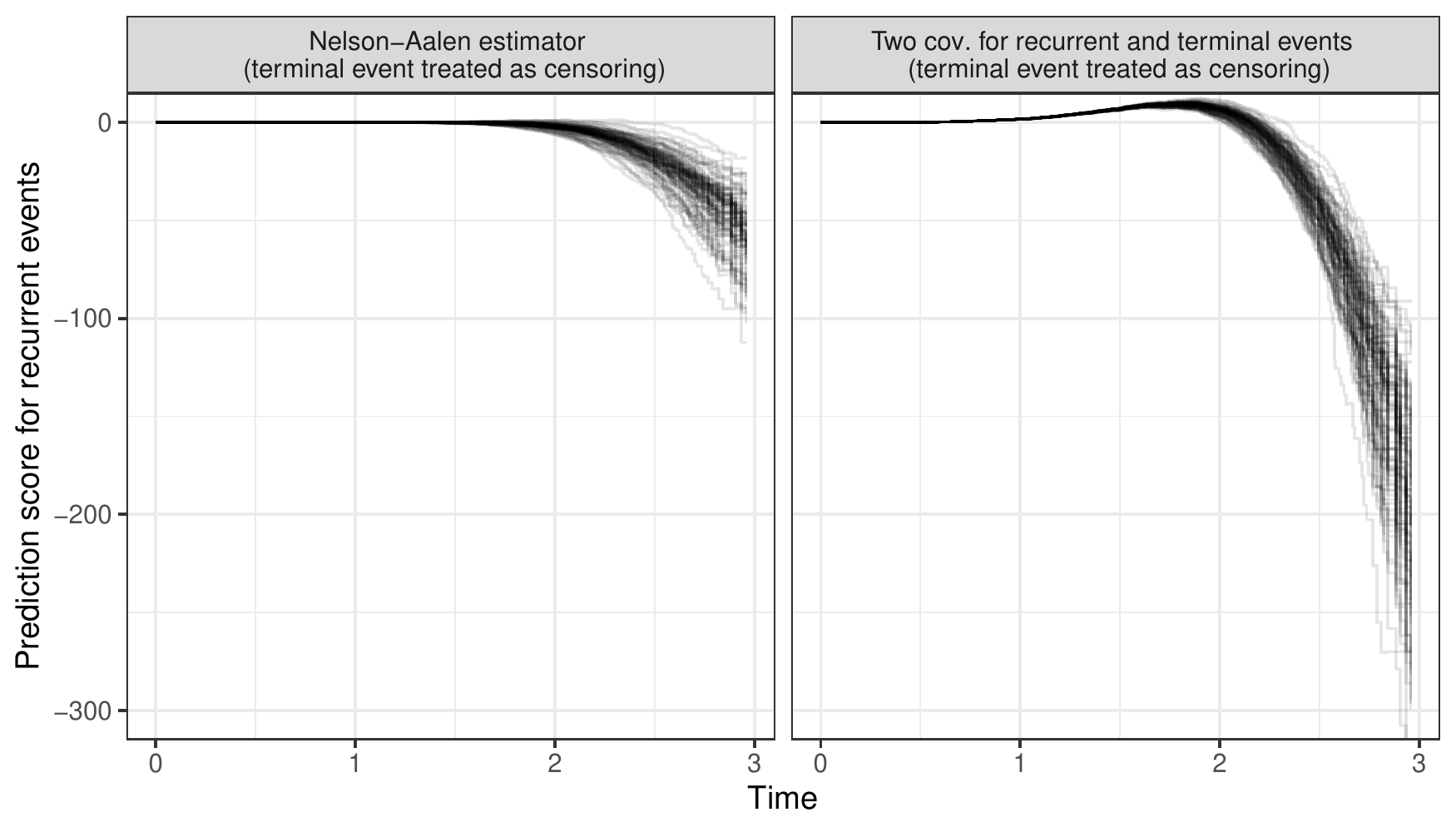}
\end{tabular}}
\caption{{\footnotesize Prediction scores for the expected cumulative number of recurrent events when the terminal event is ignored and treated as a censoring variable. The data were generated as in Figure~\ref{fig:plot3}. The expected cumulative number recurrent of events was predicted using the Cox model with one or two covariates. The reference model uses no covariates and was estimated from the Nelson-Aalen estimator from the non-parametric estimator (see Equation~\eqref{eq:ex_term_0}).}}
\label{fig:plot4}
\end{figure}

In Figure~\ref{fig:plot3}, we simulated $100$ training samples each of size $800$ and we evaluated the prediction score on a unique test sample of size $1,000$. In the bottom panel, we see that including the two covariates in the survival model increases the prediction performance as compared to the model with only one covariate. Also, for both models, the gain in terms of prediction is more important for small time points and is reduced after time $2$ approximately, as compared to the Kaplan-Meier estimator. This loss in terms of prediction efficiency of the survival function for large time points impacts the prediction of the expected cumulative number of recurrent events. In the top panel, we see that adding the correct covariates in the Cox models of the survival function and of the recurrent event models increases the prediction performances. After time $2$, the gain in terms of the performance prediction of the expected cumulative number of recurrent events slightly decreases due to the loss of efficiency in the prediction of the survival function. Table~\ref{tab:2} provides the mean score of those three different models based on $500$ training samples of size $100$, $200$, $400$ and $800$ and one single test sample of size $1,000$. We see the same trend as in Figure~\ref{fig:plot3} for all sample sizes. Clearly, increasing the sample size does not provide much gain in terms of average especially for small time points but it does reduce the variability of the predictors. 
In Figure~\ref{fig:plot4}, we also looked at models where the terminal event is treated as a censored variable. One model does not include any covariate and is estimated using the Nelson-Aalen estimator. The other includes the two correct covariates and is based on the Cox model. The reference model is the same as previously (see Equation~\eqref{eq:ex_term_0}), that is with no covariates but with the terminal event properly taken into account. We clearly see that both models provide very poor predictions. This is due to the fact that when the terminal event is treated as a censoring variable, the corresponding Nelson-Aalen estimator predicts too many recurrences over time.  


\captionsetup{font={small}}

\begin{table}[!ht]
\centering
\resizebox{\textwidth}{!}{
\begin{tabular}{|l|ccc|ccc|}
  \hline
  & \multicolumn{3}{c|}{$n_{\text{train}}=100$} & \multicolumn{3}{c|}{$n_{\text{train}}=200$}\\
 & $t=1$ & $t=2$ & $t=2.9$ & $t=1$ & $t=2$ & $t=2.9$ \\
 \hline
One cov.-one cov. & $0.68$ $(0.05)$ & $6.47$ $(0.43)$ & $6.14$ $(0.83)$ & $0.69$ $(0.03)$ & $6.54$ $(0.24)$ & $6.33$ $(0.45)$ \\ 
  Two cov.-one cov. & $1.67$ $(0.07)$ & $12$ $(0.92)$ & $6.43$ $(2.71)$ & $1.67$ $(0.04)$ & $12.11$ $(0.6)$ & $6.84$ $(1.77)$ \\ 
  Two cov.-two cov. & $1.67$ $(0.07)$ & $12.61$ $(0.59)$ & $10.07$ $(1.18)$ & $1.68$ $(0.04)$ & $12.75$ $(0.3)$ & $10.51$ $(0.59)$ \\ 
   \hline
     & \multicolumn{3}{c|}{$n_{\text{train}}=400$} & \multicolumn{3}{c|}{$n_{\text{train}}=800$}\\
 & $t=1$ & $t=2$ & $t=2.9$ & $t=1$ & $t=2$ & $t=2.9$ \\
 \hline
One cov.-one cov. & $0.69$ $(0.01)$ & $6.57$ $(0.14)$ & $6.42$ $(0.28)$ & $0.69$ $(0.01)$ & $6.59$ $(0.08)$ & $6.47$ $(0.17)$ \\ 
  Two cov.-one cov. & $1.68$ $(0.02)$ & $12.11$ $(0.39)$ & $6.92$ $(1.22)$ & $1.68$ $(0.01)$ & $12.15$ $(0.27)$ & $7.07$ $(0.85)$ \\ 
  Two cov.-two cov. & $1.68$ $(0.02)$ & $12.82$ $(0.17)$ & $10.69$ $(0.35)$ & $1.68$ $(0.01)$ & $12.84$ $(0.09)$ & $10.78$ $(0.23)$ \\ 
   \hline
\end{tabular}}
\caption{{\footnotesize Means and standard deviations (in bracket) 
over $500$ simulations for the prediction score of the expected number of recurrent events in the presence of a terminal event. The results are presented for the same three models as in Figure~\ref{fig:plot3}. The reference model uses no covariates and was estimated from the non-parametric estimator in Equation~\eqref{eq:ex_term_0}. Large values indicate better predictive performances.}} 
\label{tab:2}
\end{table}

\begin{table}[!ht]
\centering
\resizebox{\textwidth}{!}{
\begin{tabular}{|l|ccc|ccc|}
  \hline
$\times 10^4$   & \multicolumn{3}{c|}{$n_{\text{train}}=100$} & \multicolumn{3}{c|}{$n_{\text{train}}=200$}\\
 & $t=1$ & $t=2$ & $t=2.9$ & $t=1$ & $t=2$ & $t=2.9$ \\
 \hline
One cov. & $0.58$ $(0.55)$ & $77.22$ $(34.77)$ & $22.57$ $(8.29)$ & $0.69$ $(0.25)$ & $84.71$ $(15.22)$ & $23.96$ $(3.81)$ \\ 
  Two cov. & $1.71$ $(1.49)$ & $214.84$ $(46.73)$ & $32.3$ $(18.05)$ & $2.07$ $(0.73)$ & $227.83$ $(23.1)$ & $38.3$ $(8.86)$ \\ 
   \hline
  & \multicolumn{3}{c|}{$n_{\text{train}}=400$} & \multicolumn{3}{c|}{$n_{\text{train}}=800$}\\
 & $t=1$ & $t=2$ & $t=2.9$ & $t=1$ & $t=2$ & $t=2.9$ \\
 \hline
One cov. & $0.74$ $(0.17)$ & $86.26$ $(8.65)$ & $24.29$ $(2.81)$ & $0.77$ $(0.12)$ & $89.74$ $(5.44)$ & $27.22$ $(3.02)$ \\ 
  Two cov. & $2.2$ $(0.51)$ & $231.51$ $(13.94)$ & $40.49$ $(3.74)$ & $2.28$ $(0.36)$ & $238.41$ $(9.88)$ & $46.23$ $(4.27)$ \\ 
   \hline
\end{tabular}}
\caption{{\footnotesize Means and standard deviations (in bracket) 
over $500$ simulations for the prediction score of the survival function. The results were multiplied by $10^4$ and are presented for the model with one or two covariates. The reference model uses no covariates and is estimated from the Kaplan-Meier estimator. Large values indicate better predictive performance.}} 
\label{tab:3}
\end{table}

\newpage

\section{Real data analysis: the Atrial Fibrillation dataset}\label{sec:AFdata}

In this section, we analyse a dataset on patients with atrial fibrillation (AF). The aim is to compare different regression models for the prediction of the expected cumulative number of atrial fibrillation hospitalisations, using the prediction score developed in this work. Patients were enrolled from January 1st 2008 to December 1st 2012  in the ``Atrial Fibrillation Survey–Copenhagen (ATLAS-CPH)'' from both the in- and outpatient clinics at the Department of Cardiology at University Hospital Copenhagen, Hvidovre, Denmark. All patients were previously diagnosed with AF and were categorised at baseline, into either suffering from paroxysmal atrial fibrillation (PAF) or persistent atrial fibrillation (PeAF). PAF was defined as at least one recorded AF episode with spontaneous conversion to sinus rhythm, no valvular AF, and excluding other temporal forms of AF. PeAF was defined as at least one recorded episode of AF lasting more than 7 days, or where either medical or electrical cardioversion was needed to restore sinus rhythm (in accordance with the Danish Cardiology Society AF guidelines at this time). Other inclusion criteria were age $> 18$ years, recent ($< 1$ month) AF documented via either standard 12-lead electrocardiogram (ECG) or home monitoring and ability to give oral and written consent. In total, $174$ patients were enrolled with $50$ PAF patients and $124$ PeAF patients. Time is measured in days, with a mean follow-up duration of $1\,279$ days. In terms of observed events, the patients experienced a total of $325$ AF hospitalisations, with $305$ AF hospitalisations in the PeAF group and $20$ in the PAF group. A terminal event was defined as either progression to permanent AF or as the occurence of death. In the dataset, $45$ patients experienced a terminal event and the remaining $129$ patients were censored. Finally, in top of the AF type, the dataset also includes $11$ additional variables: gender, age, alcohol consumption (with two levels $0-5$ and $>5$), tobacco consumption (with three levels ``never smoked'', ``ex-smoker'', ``current smoker''), presence of hypertension, heart failure, valvular heart disease, ischemic heart disease, diabetes, COPD, antiarrythmic medication. The data are presented in great details in~\cite{schroder2019recurrent}. Note also that the data are fully available from the Plos One website.

In~\cite{schroder2019recurrent}, the authors analysed the data using a multi-state approach with four possible states: no experience of recurrent events yet, $1$ recurrent event, $2$ or more recurrent events and the absorbing state for the terminal event. The transition intensities were assumed to be proportional with each other using a Cox model, where the number of previous recurrent events was included in the model. Those types of multi-state models with terminal event are described for instance in~\cite{cook2007statistical} (see Section 6.6.4 of their book). Those analyses showed a high significant effect of the AF type, the number of previous recurrent events (p-values $<10^{-4}$) and of age (p-value$=0.0253$) for the risk of future AF hospitalisations. The effect of diabetes had a p-value equal to $0.0955$. All other variables were assessed as non significants (p-values$>0.2$). A Cox model was also implemented for the terminal event using the multi-state approach (that is including the effect of previous AF hospitalisations through a proportional effect) with all variables. Only the age variable was significant (in the multivariate Cox model, the hazard rate was equal to $1.05$ and the p-value was equal to $0.0016$). In this previous work, the authors then decided to only include the covariates AF type and age, with a proportional effect of the number of previous AF hospitalisations for the modelisation of the recurrent event process. For the terminal event model, they only included the age variable. Based on those models, it is then possible to produce predictions for the expected cumulative number of future AF hospitalisations, for a given patient based on his/her characteristics. Using the prediction score developed in this paper, we will compare the performance of the model used in~\cite{schroder2019recurrent} with several other possible models. 
Since diabetes is a known risk factor for AF, we will also consider models with this variable, along with AF type and age. As in the simulation section, the prediction score will be computed from a training and a test samples, but this time using $10$-fold cross validation, that is one tenth of the observations are used for the test sample and the remaining observations are used for the model estimations and the procedure is repeated and averaged ten times.

\begin{figure}[!p]
\centering
\resizebox{\textwidth}{!}{ 
\begin{tabular}{c}
\includegraphics[width=0.1\textwidth]{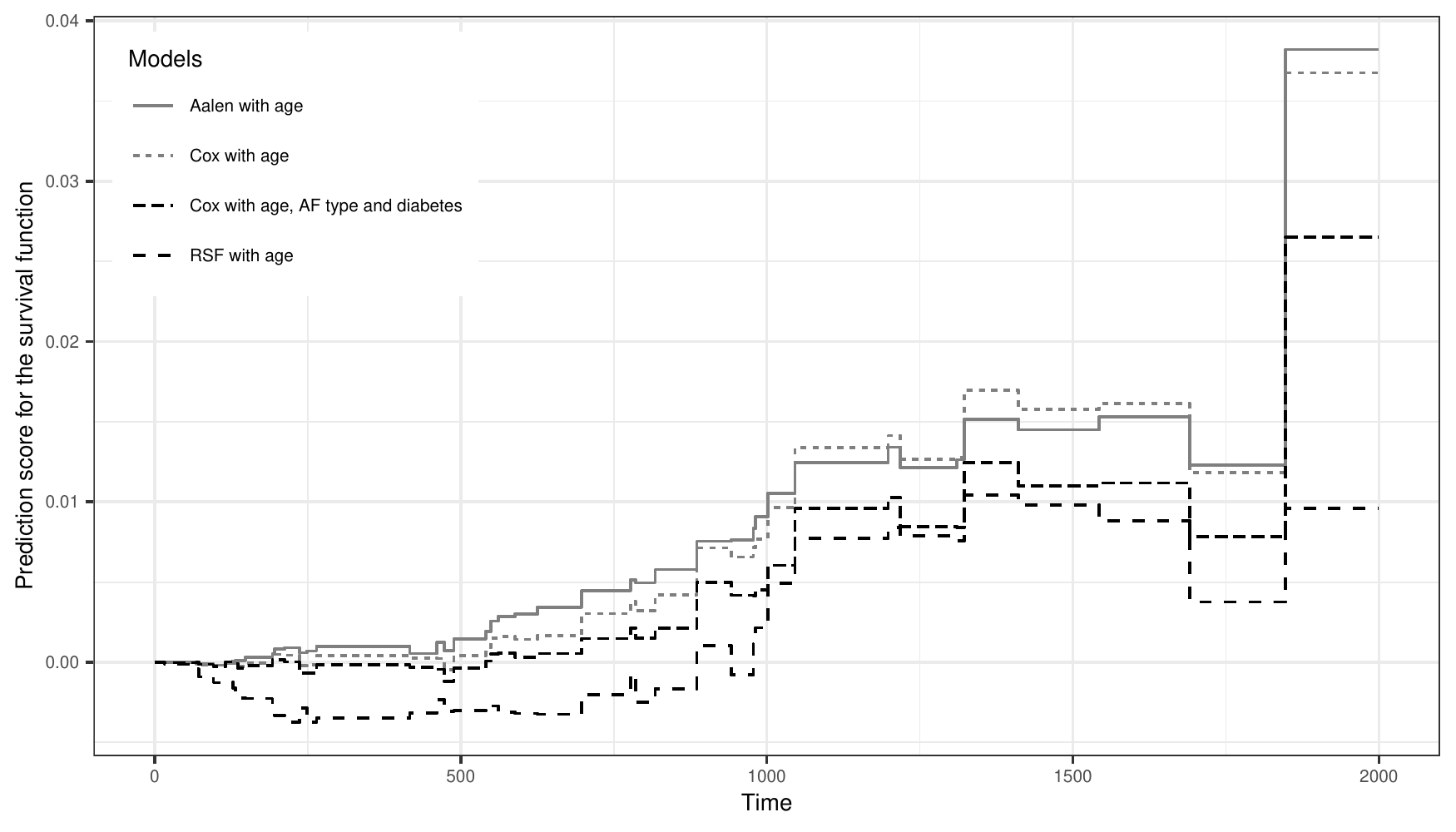}
\end{tabular}}
\caption{{\footnotesize Prediction scores for the survival function of the terminal event in the atrial fibrillation dataset, computed using $10$-fold cross validation. With the Kaplan-Meier estimator as the reference, four different models are compared. For ease of visualisation, we describe those models in increasing order of their scores at time $t=2000$: the Cox model with covariates age, AF type and diabetes (score $=0.067$), the Random Survival Forest (RSF) model with covariate age (score $=0.069$),  the Aalen model with covariate age (score $=0.094$) and the Cox model with covariate age (score $=0.105$).}}
\label{fig:plot_Atrial1}
\end{figure}

\begin{figure}[!p]
\centering
\resizebox{\textwidth}{!}{ 
\begin{tabular}{c}
\includegraphics[width=0.1\textwidth]{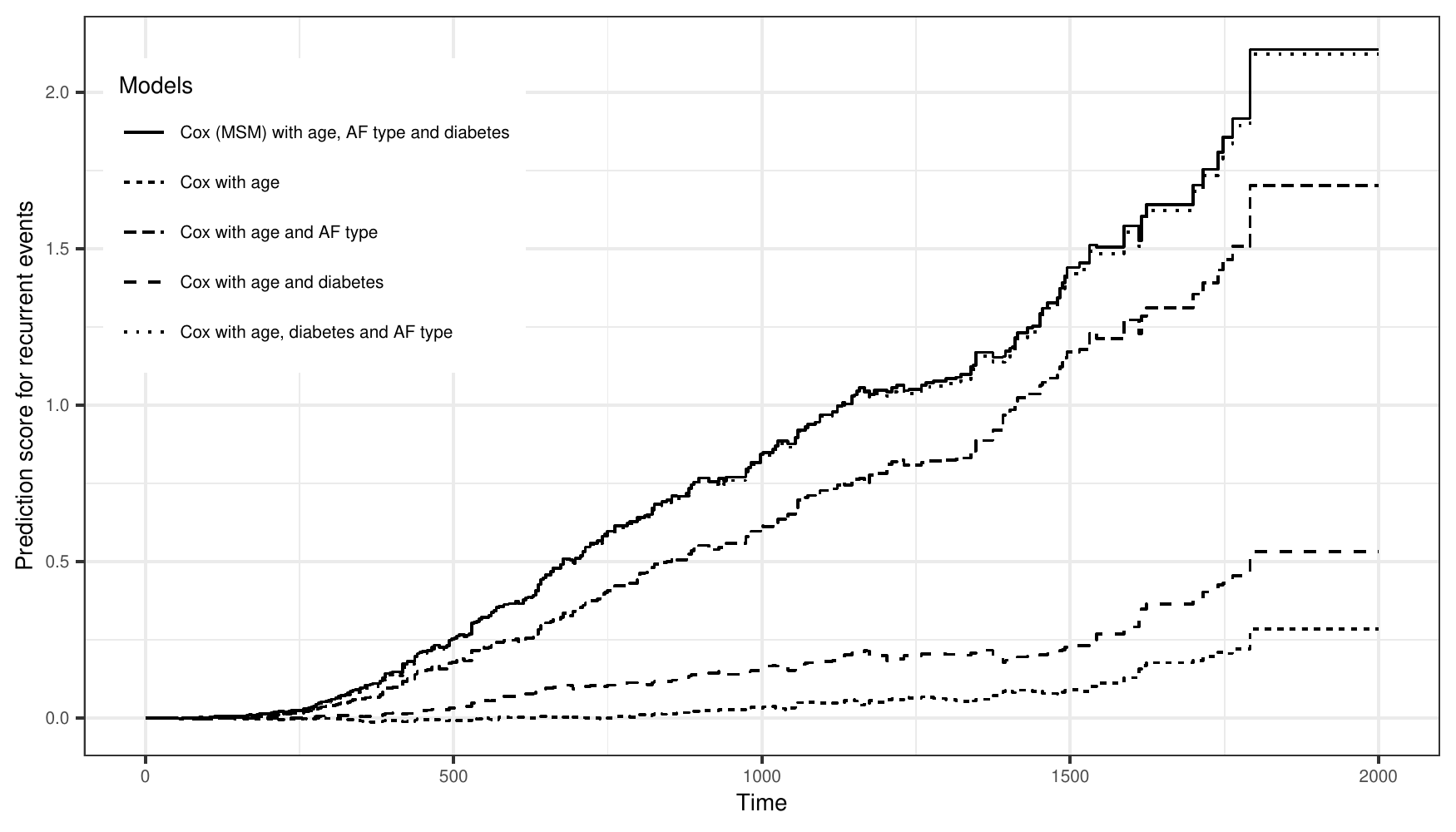}
\end{tabular}}
\caption{{\footnotesize Prediction scores for the expected cumulative number of recurrent events in the atrial fibrillation dataset. With the non-parametric estimator (see Equation~\eqref{eq:ex_term_0}) as the reference, six different models are compared. All the models use the Cox model with age as the unique covariate for the estimation of the survival function. For ease of visualisation, we describe the six models (for the recurrent events) in increasing order of their scores at time $t=2000$: the Cox model with covariate age (score $=0.911$), the Cox model with covariates age and diabetes (score $=1.572$), the Cox model with covariate age and AF type (score $=4.340$), the multi-state (MSM) Cox model with covariates age, AF type and diabetes (score $=5.241$) and the Cox model with covariates age, AF type and diabetes (score $=5.431$). The MSM Cox model assumes that the transition intensities from $0$ event to $1$ and to ``one event or more'' to a new event are proportionals.}}
\label{fig:plot_Atrial2}
\end{figure}

\begin{figure}[!p]
\centering
\resizebox{\textwidth}{!}{ 
\begin{tabular}{c}
\includegraphics[width=0.1\textwidth]{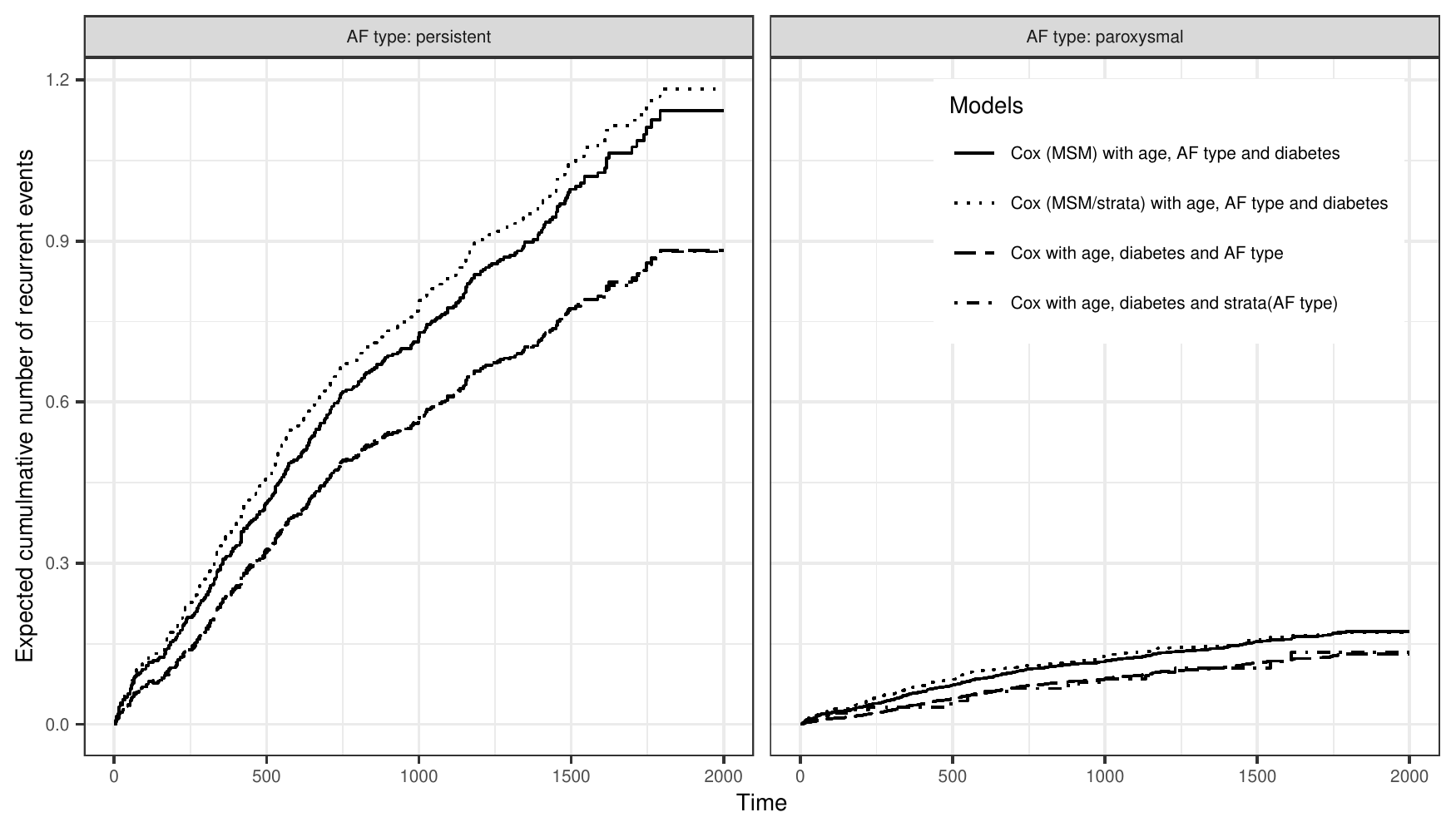}
\end{tabular}}
\caption{{\footnotesize Expected cumulative number of recurrent events predicted from four different models for two $60$ year old patients with diabetes, one with persistent AF and the other with paroxysmal AF. Those models had similar prediction performances (see Table~\ref{tab:AF_rec} and Figure~\ref{fig:plot_Atrial2}).}}
\label{fig:plot_Atrial3}
\end{figure}

We first evaluate the prediction performance for the terminal event with the Cox models with age only and with age, AF type and diabetes, the Aalen model with age and the random survival forests with age. The reference model is taken as the Kaplan-Meier estimator and the score is computed using formula~\eqref{eq:score}, where the prediction criterion $\widehat{\mathrm{MSE}}$ is computed from the Kaplan-Meier estimator of the censoring variable. The random survival forests were implemented from the \texttt{rfsrc} package (see~\cite{ishwaran2008random}). The results are presented in Table~\ref{tab:AF_surv} and Figure~\ref{fig:plot_Atrial1}. We clearly see that the survival random forests perform poorly, especially before time $1\,000$ where the Kaplan-Meier shows a better performance. The Cox model with age, AF type and diabetes shows a better performance for all time points and the Aalen and Cox models with age show very similar performances and outperform all four models. Other models were also investigated with the different combinations of all three variables with each algorithm and the results were similar and are therefore omitted. In the following, we now decide to use the Cox model with age for the modelling of the terminal event. The prediction performance for the recurrent event process is now investigated. We consider the following models: four multivariate Cox models based on the age, diabetes and AF type variables, the stratified Cox model with respect to AF type with the variables age and diabetes, the Aalen model with all three variables, the Cox multi-state model with the three variables that was used in~\cite{schroder2019recurrent} and the same Cox multi-state model but stratified with respect to AF type. The reference model is taken as the non-parametric estimator (see Equation~\eqref{eq:ex_term_0}) and the score is again computed using formula~\eqref{eq:score}. The results are displayed in Table~\ref{tab:AF_rec} and Figure~\ref{fig:plot_Atrial2} (in the figure only five different models are represented). We observe that the Cox model with the age variable has a poor predictive performance. From this model, adding the diabetes or AF type improves the model, with a much bigger gain with the AF type variable. Further, combining all three variables in the same model provides a substantial gain with respect to all previous models. On the other hand, the two multi-state models provide only a minor improvement of the predictions with a slight advantage for the stratified model. Finally, the predictions for some of these models on the expected cumulative number of AF hospitalisations are displayed in Figure~\ref{fig:plot_Atrial3}. In this figure, the predictions are made for two $60$ year old patients with diabetes, one with persistent AF and the other with paroxysmal AF. While the different models does not vary much in their predictions for the paroxysmal AF patient, they offer different results for the persistent AF patient. According to the results from Table~\ref{tab:AF_rec} and Figure~\ref{fig:plot_Atrial2}, the Cox multi-state model stratified with respect to AF type has the greatest prediction performance and therefore should be chosen. After $1\,500$ days after AF diagnosis, this model predicts an expected number of AF hospitalisations equal to $1.03$ approximately. On the other hand, if one uses the multi-state Cox model, the prediction is equal to $0.98$, if one uses the multivariate Cox model with all three variables, the prediction is equal to $0.76$.



\captionsetup{font={small}}

%

\begin{table}[!p]
\centering
\resizebox{\textwidth}{!}{ 
\begin{tabular}{|l|c|c|c|}
  \hline
  & $t=1000$ & $t=1500$ & $t=2000$ \\
 \hline
Aalen with age & $0.011\; [0.002,0.017]$ & $0.015\; [0.001,0.039]$ & $0.038\; [-0.001,0.035]$ \\ 
  Cox with age & $0.010\; [-0.006,0.023]$ & $0.016\; [-0.001,0.042]$ & $0.037\; [-0.005,0.053]$ \\ 
  Cox with age, AF type, diabetes & $0.006\; [-0.009,0.019]$ & $0.011\; [-0.007,0.04]$ & $0.027\; [-0.002,0.04]$ \\ 
  RSF with age & $0.005\; [-0.021,0.032]$ & $0.010\; [-0.014,0.039]$ & $0.010\; [-0.031,0.045]$ \\ 
   \hline
\end{tabular}}
\caption{{\footnotesize Means and $80\%$ intervals (in curly bracket) 
over $10$-folds cross validation for the prediction score of the survival function of the terminal 
event in the atrial fibrillation dataset. With the Kaplan-Meier estimator as the reference, four different models are compared 
at three different time points: the Aalen and Cox models with covariate age, the Cox model with covariates 
age, AF type and diabetes and the Random Survival Forest model with covariate age.}} 
\label{tab:AF_surv}
\end{table}


\begin{table}[!p]
\centering
\resizebox{\textwidth}{!}{
\begin{tabular}{|l|c|c|c|}
  \hline
  & $t=1000$ & $t=1500$ & $t=2000$ \\
 \hline
Cox with age & $0.036\; [-0.248,0.289]$ & $0.089\; [-0.690,0.629]$ & $0.283\; [-0.719,0.985]$ \\ 
  Cox with age and diabetes & $0.166\; [-0.342,0.538]$ & $0.232\; [-0.372,1.051]$ & $0.532\; [-0.640,1.495]$ \\ 
  Cox with age and AF type & $0.611\; [0.104,1.138]$ & $1.170\; [0.278,2.081]$ & $1.702\; [-0.027,3.989]$ \\ 
  Cox with age, AF type and diabetes & $0.840\; [-0.123,1.539]$ & $1.420\; [-0.033,2.983]$ & $2.121\; [-0.485,5.683]$ \\ 
  Aalen with age, AF type and diabetes & $0.818\; [0.143,1.312]$ & $1.482\; [0.290,2.992]$ & $2.140\; [-0.190,5.506]$ \\ 
  Cox with age, diabetes and strata(AF type) & $0.836\; [-0.133,1.521]$ & $1.416\; [-0.053,3.000]$ & $2.118\; [-0.468,5.667]$ \\ 
  Cox (MSM) with age, AF type and diabetes & $0.847\; [-0.041,1.459]$ & $1.439\; [0.084,2.893]$ & $2.136\; [-0.310,5.526]$ \\ 
  Cox (MSM/strata) with age, AF type and diabetes & $0.852\; [-0.014,1.475]$ & $1.454\; [0.108,3.073]$ & $2.147\; [-0.298,5.756]$ \\ 
   \hline
\end{tabular}}
\caption{{\footnotesize Means and $80\%$ intervals (in curly bracket) 
over $10$-folds cross validation for the prediction score of the expected number of recurrent 
events in the atrial fibrillation dataset. With the non-parametric estimator (see Equation~\eqref{eq:ex_term_0}) 
as the reference, eight different models (for the recurrent events) are compared at three different time points: five Cox and Aalen models
with covariates age, diabetes and AF type, one Cox model stratified with respect to AF type and two multi-state 
Cox models,  Cox (MSM) and Cox (MSM/strata). The 
difference between the last two models is that the first one assumes the baseline transition intensities from $0$ event to $1$ and 
to ``one event or more'' to a new event to be proportionals while the last one uses two different baselines functions. 
All the models use the Cox model with age as the unique covariate for the estimation of the 
survival function.}} 
\label{tab:AF_rec}
\end{table}

\captionsetup{font={normalsize}}

\newpage

\section{Conclusion}

In this work a new prediction criterion was proposed in the context of recurrent event data. The criterion evaluates the prediction performance of the expected cumulative number of recurrent events, while taking into account censoring and a possible terminal event. We showed that it can be decomposed into an inseparability and imprecision terms in the same manner as in~\cite{graf1999assessment}. Moreover, the simulations revealed that the inseparability term was largely dominant in the decomposition. As a result, we recommend to use the prediction score defined in Equation~\eqref{eq:score}, as the difference between the prediction criterion of a given model and of a reference model, typically a model that does make use of the covariates, such that the score provides the absolute gain from the covariates in the proposed model. An alternative score could be derived by computing the relative gain as proposed in~\cite{steyerberg2010assessing}. This produces a score that ranges from $0\%$ to $100\%$ and shares similarities with the Pearson's $\mathrm{R}^2$ statistic. However, care should be taken with such a score, due to the fact that we normalise with respect to the prediction criterion of the reference model, which itself can be decomposed into imprecision and inseparability. This criterion could therefore be misleading due to the magnitude of the inseparability term which is unknown in practice.

The proposed prediction criterion is simple to compute and has the advantage to include all the recurrent events. As a result, it can be seen as an overall performance measure that provides information about the global predictive ability of the proposed model in a recurrent event context. Nevertheless, it would be possible to modify the criterion if one is interested into evaluating the performance of a model to only predict further recurrent events after a fixed number of events have already been experienced by a patient. This would amount to condition on a given number of experienced recurrent events in a multi-state framework. This type of criterion would be similar to the one developed in~\cite{schoop2011measures} which conditions on being alive up to a time $t^*$ and evaluate the prediction of the model for a time $s>t^*$. Another improvement would be to allow for frailty models in the manner of~\cite{van2016exploring}. A marginal score that integrates the frailty variable could be derived. Such a score would provide an overall evaluation of the frailty model and would be a natural extension of the score proposed in this paper. Alternatively, a conditional score could be proposed for the conditional (with respect to the frailty) expected cumulative number of recurrent events. More work is needed to develop these two scores.
  
 

\section{Appendix: proofs of the convergence of the prediction criterion for the expected cumulative number of recurrent events under the two scenarios}\label{sec:proof}


In the proof of Proposition~\ref{th:0} we will use a different decomposition of the observed recurrent event process $N$, depending on the scenario under study. With a slight abuse of notations we use the same name, $M$, for the residual process in both scenarios.
\begin{itemize}
\item In the case of right-censoring only (Section~\ref{sec:RC}) we have:
\begin{align*}
N(t)-\int_0^t I(C\geq u) \lambda^*(u\mid X(u))du=M(t),
\end{align*}
with $\mathbb E[dM(t)\mid \{I(C\geq u),X(u) : 0\leq u\}]=0$.
\item In the case of right-censoring and terminal event (Section~\ref{sec:terminal}) we have:
\begin{align*}
N(t)-\int_0^t I(T\geq u) \lambda^*(u\mid X(u))du=M(t),
\end{align*}
with $\mathbb E[dM(t)\mid \{I(T\geq u),X(u) : 0\leq u\}]=0$.
\end{itemize}

\subsection{Proof of Proposition~\ref{th:0}}\label{proof:th0}

In both scenarios, we directly have:
\begin{align*}
\mathrm{MSE}(t,\mu) &= \mathbb E\left[\big(\mu(t\mid X(t))-\mu^*(t\mid X(t))\big)^2\right] + \mathbb E\left[\Bigg(\int_0^t\frac{dN(u)}{1-G(u-\mid X(u))}-\mu^*(t\mid X(t))\Bigg)^2\right]\\
&\quad + 2 \mathbb E\left[\Bigg(\int_0^t\frac{dN(u)}{1-G(u-\mid X(u))}-\mu^*(t\mid X(t))\Bigg)\Bigg(\mu^*(t\mid X(t))-\mu(t\mid X(t))\Bigg)\right].
\end{align*}
Using the fact that $\mathbb E[\int_0^tdN(u)/(1-G(u-\mid X(u)))\mid X(t)]=\mu^*(t\mid X(t))$, we conclude that
\begin{align*}
\mathrm{MSE}(t,\mu) &=\mathbb E\left[(\mu(t\mid X(t))-\mu^*(t\mid X(t)))^2\right]+A(t),
\end{align*}
where
\begin{align}\label{eq:A}
A(t)&=\mathbb E\left[\left(\int_0^t\frac{dN(u)}{1-G(u-\mid X(u))}\right)^2\right]-\mathbb E\left[\left(\mu^*(t\mid X(t))\right)^2 \right].
\end{align}
We now show, in the case of right-censoring only, that $A(t)\geq 0$. Using the decomposition given at the beginning of Section~\ref{sec:proof}, we have,
\begin{align*}
\mathbb E\left[\left(\int_0^t\frac{dN(u)}{1-G(u-\mid X(u))}\right)^2\right]&=\mathbb E\left[\left(\int_0^t\frac{dM(u)}{1-G(u-\mid X(u))}\right)^2\right]\\
&\quad+\mathbb E\left[\left(\int_0^t\frac{\lambda^*(u\mid X(u))I(C\geq u)du}{1-G(u-\mid X(u))}\right)^2\right],
\end{align*}
since 
\begin{align*}
& \mathbb E\left[\int_0^t\frac{dM(u)}{1-G(u-\mid X(u))}\int_0^t\frac{\lambda^*(v\mid X(v))I(C\geq v)dv}{1-G(v-\mid X(v))}\right]\\
& =\mathbb E\left[\int_0^t\frac{\mathbb E[dM(u)\mid \{I(C\geq v),X(v) : 0\leq v\}}{1-G(u-\mid X(u))}\int_0^t\frac{\lambda^*(v\mid X(v))I(C\geq v)dv}{1-G(v-\mid X(v))}\right]=0.
\end{align*}
Then,
\begin{align*}
\mathbb E\left[\left(\int_0^t\frac{\lambda^*(u\mid X(u))I(C\geq u)du}{1-G(u-\mid X(u))}\right)^2\right] & =2\mathbb E\left[\int_{0\leq u< v\leq t}\frac{\lambda^*(u\mid X(u))\lambda^*(v\mid X(v))}{1-G(u-\mid X(u))}dudv\right]\\
\mathbb E\left[\left(\mu^*(t\mid X(t))\right)^2 \right]& =2\mathbb E\left[\int_{0\leq u< v\leq t}\lambda^*(u\mid X(u))\lambda^*(v\mid X(v))dudv\right].
\end{align*}
As a result,
\begin{align*}
A(t) & =\mathbb E\left[\left(\int_0^t\frac{dM(u)}{1-G(u-\mid X(u))}\right)^2\right]\\
&\quad+2\mathbb E\left[\left(\int_{0\leq u< v\leq t}\frac{\lambda^*(u\mid X(u))\lambda^*(v\mid X(v))G(u-\mid X(u))}{1-G(u-\mid X(u))}dudv\right)\right]\geq 0.
\end{align*}
The proof in the presence of a terminal event is similar. We find 
\begin{align*}
A(t) & =\mathbb E\left[\left(\int_0^t\frac{dM(u)}{1-G(u-\mid X(u))}\right)^2\right]\\
&\quad+2\mathbb E\left[\left(\int_{0\leq u< v\leq t}\frac{1-S(u\mid X(u))(1-G(u-\mid X(u)))}{1-G(u-\mid X(u))}\right.\right.\\
&\qquad \;\;\;\;\;\;\;\;\;\; \times\lambda^*(u\mid X(u))\lambda^*(v\mid X(v))S(v\mid X(v))dudv\bigg)\bigg]\geq 0.
\end{align*}

\subsection{Proof of Proposition~\ref{th:1}}\label{proof:th1}
We start by proving that $\mathbb E[\mu^*(\tau\mid X(\tau))]<\infty$ in the presence of a terminal event (the scenario without terminal event follows from the same arguments). 
We have for all $t\in[0,\tau]: \mathbb P[C\geq t\mid X(t)] \geq \mathbb P[T\geq t\mid X(t)]\geq c$, from Assumption~\ref{as:2}. From the same assumption, $N(\tau)$ is almost surely bounded by a constant. As a consequence, 
\begin{align*}
\mu^*(\tau\mid X(\tau))=\int_0^{\tau} \frac{\mathbb E[dN(t)\mid X(t)]}{1-G(t-\mid X(t))}
\end{align*}
is almost surely bounded, where the equality has been proved in Section~\ref{sec:terminal}. The rest of the proof of Proposition~\ref{th:1} is identical in both scenarios. 

We first note $F_{X(t)}(x)=\mathbb P[X(t)\leq x]$, we let $\mathcal X_{u,v}$ denote the support of the joint distribution $(X(u),X(v))$ and we note $F_{X(u),X(v)}(x,y)=\mathbb P[X(u)\leq x,X(v)\leq v]$. We then introduce the quantity
\begin{align*}
\xi(t)&=\int_{0\leq u, v\leq t}\int_{\mathcal X_{u,v}}\frac{\mathbb E[dN(u)dN(v) \mid X(u)=x,X(v)=y]}{(1-\hat G(u-\mid x))(1-\hat G(v-\mid y))}dF_{X(u),X(v)}(x,y)\\
&\quad -2  \int_{\mathcal X_{t}}\widehat \mu(t\mid x)\mu^*(t\mid x)dF_{X(t)}(x)\\
&\quad +\int_{\mathcal X_t} \left(\widehat \mu(t\mid x)\right)^2dF_{X(t)}(x)=:\xi_1(t)+\xi_2(t)+\xi_3(t).
\end{align*}
Write:
\begin{align*}
 \left|\widehat{\mathrm{MSE}^1}(t,\hat\mu)-\mathrm{MSE}^1(t,\mu)\right| &\leq \left|\xi(t)-\mathbb E\bigg[\bigg(\int_0^t \frac{dN(u)}{1-G(u-\mid X(u))}-\mu(t\mid X(t))\bigg)^2\bigg]\right|\\
& \quad + \left|\frac 1n \sum_{i=1}^n \bigg(\int_0^t \frac{dN_i(u)}{1-\hat G(u-\mid X_i(u))}-\widehat\mu(t\mid X_i(t))\bigg)^2-\xi(t)\right|\\
&\leq : C(t)+D(t).
\end{align*}
By decomposing the square term into three other terms, we bound $C(t)$ in the following way: $C(t)\leq |C_1(t)|+|C_2(t)|+|C_3(t)|$ with
\begin{align*}
C_1(t)
& =\int_{0\leq u,v\leq t}\int_{\mathcal X_{u,v}}\frac{(1-G(u-\mid x))(1-G(v-\mid y))-(1-\hat G(u-\mid x))(1-\hat G(v-\mid y))}{(1-\hat G(u-\mid x))(1-\hat G(v-\mid y))(1-G(u-\mid x))(1-G(v-\mid y))}\\
&\qquad \mathbb E[dN(u)dN(v) \mid X(u)=x,X(v)=y]dF_{X(u),X(v)}(x,y),\\
C_2(t) & =-2  \int_{\mathcal X_{t}}(\widehat \mu(t\mid x)- \mu(t\mid x))\mu^*(t\mid x)dF_{X(t)}(x),\\
C_3(t)&=\int_{\mathcal X_t} \left(\left(\widehat \mu(t\mid x)\right)^2-\left(\mu(t\mid x)\right)^2\right)dF_{X(t)}(x).
\end{align*}

For $C_1(t)$ we have
\begin{align*}
& (1-G(u-\mid x))(1-G(v-\mid y))-(1-\hat G(u-\mid x))(1-\hat G(v-\mid y))\\
&=\left(\hat G(u-\mid x)-G(u-\mid x)\right)+\left(\hat G(v-\mid y)-G(v-\mid y)\right)\\
&\quad+G(u-\mid x)\left(G(v-\mid y)-\hat G(v-\mid y)\right)+\hat G(v-\mid y)\left(G(u-\mid x)-\hat G(u-\mid x)\right),
\end{align*}
and we can deal with all four terms in the same fashion. For instance, for the first term,
\begin{align*}
&\int_{0\leq u, v\leq t}\int_{\mathcal X_{u,v}}\frac{\left(\hat G(u-\mid x)-G(u-\mid x)\right)\mathbb E[dN(u)dN(v) \mid X(u)=x,X(v)=y]}{(1-\hat G(u-\mid x))(1-\hat G(v-\mid y))(1-G(u-\mid x))(1-G(v-\mid y))}dF_{X(u),X(v)}(x,y)\\
&\quad \leq \int_{0}^{t}\int_{\mathcal X_{u}}\frac{\left|\hat G(u-\mid x)-G(u-\mid x)\right|\mathbb E[dN(u) \mid X(u)=x]}{(1-\hat G(u-\mid x))(1-G(u-\mid x))}dF_{X(u)}(x),
\end{align*}
using the fact that $\int_0^t dN(v)/((1-\hat G(v-\mid y))(1-G(v-\mid y)))$ is bounded. Then, since $\int_0^t\mathbb E[dN(u)/(1-G(u-\mid X(u))) \mid X(u)=x]=\mu^*(t\mid X(t))$ 
and $(1-\hat G(u-\mid x))^{-1}$ is asymptotically uniformly bounded, we conclude that $|C_1(t)|$ tends toward $0$ in probability using the uniform consistency of the censoring estimator.

For $C_2(t)$ we use the consistency of $\widehat \mu$ and the fact that $\mathbb E[\mu^*(t\mid X(t))]$ is finite to prove that 
$|C_2(t)|$ tends towards $0$ in probability.

For $C_3(t)$, we directly write $(\widehat \mu(t\mid x))^2-(\mu(t\mid x))^2=(\widehat \mu(t\mid x)-\mu(t\mid x))(\widehat \mu(t\mid x)+\mu(t\mid x))$ and we use the fact that $\mu(t\mid x)$ is bounded and the consistency of $\widehat \mu$ to prove that $|C_3(t)|$ tends towards $0$ in probability.

Similarly to $C(t)$ we obtain the following bound: $D(t)\leq |D_1(t)|+|D_2(t)|+|D_3(t)|$ with 
\begin{align*}
D_1(t) &=\frac 1n \sum_{i=1}^n \int_{0\leq u, v\leq t} \frac{dN_i(u)dN_i(v)}{(1-\hat G(u-\mid X_i(u)))(1-\hat G(v-\mid X_i(v)))}-\xi_1(t),\\
D_2(t) &=-\frac 2n \sum_{i=1}^n\int_0^t \frac{dN_i(u)}{1-\hat G(u-\mid X_i(u))}\widehat\mu(t\mid X_i(t))-\xi_2(t),\\
D_3(t) &=\frac 1n \sum_{i=1}^n \Big(\widehat\mu(t\mid X_i(t))\Big)^2-\xi_3(t).
\end{align*}
We now use the bound $|D_1(t)|\leq |D_{1,1}(t)| + |D_{1,2}(t)|+|D_{1,3}(t)|$ with
\begin{align*}
D_{1,1}(t)&=\frac 1n \sum_{i=1}^n \int_{0\leq u,v\leq t} \frac{dN_i(u)dN_i(v)}{(1-G(u-\mid X_i(u)))(1-G(v-\mid X_i(v)))}\\
& \quad - \int_{0\leq u, v\leq t}\int_{\mathcal X_{u,v}}\frac{\mathbb E[dN(u)dN(v) \mid X(u)=x,X(v)=y]}{(1-G(u-\mid x))(1-G(v-\mid y))}dF_{X(u),X(v)}(x,y),\\
D_{1,2}(t)&=\frac 1n \sum_{i=1}^n \int_{0\leq u,v\leq t}\chi(u,v,X_i(u),X_i(v))dN_i(u)dN_i(v)\\
D_{1,3}(t)&=- \int_{0\leq u, v\leq t}\int_{\mathcal X_{u,v}}\chi(u,v,x,y)\mathbb E[dN(u)dN(v) \mid X(u)=x,X(v)=y]dF_{X(u),X(v)}(x,y)
\end{align*}
and
\begin{align*}
\chi(u,v,x,y)&=\left\{\left(\hat G(u-\mid x)-G(u-\mid x)\right)+\left(\hat G(v-\mid y)-G(v-\mid y)\right)\right.\\
&\quad+\left(\hat G(v-\mid y)-G(v-\mid y)\right)+G(u-\mid x)\left(G(v-\mid y)-\hat G(v-\mid y)\right)\\
&\quad +\left.\hat G(v-\mid x)\left(G(u-\mid y)-\hat G(u-\mid y)\right)\right\}\\
&\quad \times \frac{1}{(1-\hat G(u-\mid x))(1-\hat G(v-\mid y))(1-G(u-\mid x))(1-G(v-\mid y))}\cdot
\end{align*}
The term $|D_{1,1}(t)|$ converges towards $0$ in probability from the strong law of large numbers. The term $|D_{1,2}(t)|$ is bounded by 
\begin{align*}
\sup_{u,v,x,y} |\chi(u,v,x,y)| \frac 1n \sum_{i=1}^n \int_{0\leq u< v\leq t}dN_i(u)dN_i(v),
\end{align*}
$\sup_{u,v,x,y} |\chi(u,v,x,y)|$ converges towards $0$ from the uniform consistency of $\hat G$ while the other term converges towards a bounded quantity from the law of large numbers. The same argument applies to $|D_{1,3}(t)|$ which also converges towards $0$ in probability.

For $D_2(t)$ we write $|D_2(t)|\leq |D_{2,1}(t)| + |D_{2,2}(t)|+ |D_{2,3}(t)|+ |D_{2,4}(t)|$ with
\begin{align*}
D_{2,1}(t)&=-\frac 2n \sum_{i=1}^n\int_0^t \frac{dN_i(u)}{1-G(u-\mid X_i(u))}\mu(t\mid X_i(t))+2  \int_{\mathcal X_{t}}\mu(t\mid x)\mu^*(t\mid x)dF_{X(t)}(x),\\
D_{2,2}(t)&=\frac 2n \sum_{i=1}^n\int_0^t \frac{dN_i(u)}{1-G(u-\mid X_i(u))}(\mu(t\mid X_i(t))-\widehat\mu(t\mid X_i(t)))\\
D_{2,3}(t)&=2  \int_{\mathcal X_{t}}(\widehat \mu(t\mid x)-\mu(t\mid x))\mu^*(t\mid x)dF_{X(t)}(x),\\
D_{2,4}(t)&=\frac 2n \sum_{i=1}^n\int_0^t \frac{(G(u-\mid X_i(u))-\hat G(u-\mid X_i(u)))dN_i(u)}{(1-G(u-\mid X_i(u)))(1-\hat G(u-\mid X_i(u)))}\widehat\mu(t\mid X_i(t)).
\end{align*}
The $D_{2,1}(t)$ term converges towards $0$ in probability from the law of large numbers. For $D_{2,2}(t)$, $D_{2,3}(t)$ and $D_{2,4}(t)$ we use the consistency of $\widehat \mu$, the convergence in probability of $ \sum_i\int_0^t dN_i(u))(1-G(u-\mid X_i(u)))/n$, the boundedness of $\mathbb E[\mu^*(t\mid X(t))]$, the uniform consistency of $\hat G$ and the asymptotic boundedness of $\widehat \mu$ and $(1-\hat G(u-\mid x))^{-1}$ to prove that all three terms converge towards $0$ in probability.

Finally, for $D_3(t)$, we write
\begin{align*}
D_3(t) & = \frac 1n \sum_{i=1}^n \Big(\mu(t\mid X_i(t))\Big)^2-\int_{\mathcal X_t}\Big(\mu(t\mid x)\Big)^2dF_{X(t)}(x)\\
& \quad + \frac 1n \sum_{i=1}^n\left(\Big(\widehat\mu(t\mid X_i(t))\Big)^2-\Big(\mu(t\mid X_i(t))\Big)^2\right)\\
& \quad +\int_{\mathcal X_t}\left(\Big(\widehat\mu(t\mid x)\Big)^2-\Big(\mu(t\mid x)\Big)^2\right)dF_{X(t)}(x).\\
\end{align*}
Each of the three terms converges towards $0$ in probability using the law of large numbers for the first term and the uniform consistency of $\widehat \mu$ for the other two.

\subsection{Proof of Proposition~\ref{th:2}}\label{proof:th3}

First, note that the Brier score can be written in the following way:
\begin{align*}
\mathrm{MSE^{Brier}}(t,\pi)= \mathbb E[S(t\mid X)] - 2\mathbb E[S(t\mid X)\pi(t\mid X)]+\mathbb E[(\pi(t\mid X))^2]. 
\end{align*}
We now study, our prediction score $\mathrm{MSE}'(t,\pi)$. Using standard martingale properties (see for instance~\cite{ABGK}), we directly have that $\mathbb E[dN(t)\mid X]=H(t\mid X)\lambda^*(t\mid X)dt$, where $H(t\mid X)=\mathbb P[T>t\mid X]=S(t\mid X) (1-G(t\mid X))$ under independent censoring and $\lambda^*$ is the hazard rate of $T^*$. As a consequence,
\begin{align}\label{eq:Brier_eq}
\mathbb E\left[\int_0^t \frac{dN(u)}{1-G(u-)}\mid X\right]=\int_0^t S(u\mid X)\lambda^*(u\mid X)du=1-S(t\mid X),
\end{align}
since $S(u\mid X)\lambda^*(u\mid X)$ is equal to the conditional density function of $T^*$. Also, it is important to notice that
\begin{align*}
\mathbb E\left[\left(\int_0^t \frac{dN(u)}{1-G(u-)}\right)^2\right]=\mathbb E\left[\int_0^t \frac{dN(u)}{(1-G(u-))^2}\right]
=\mathbb E\left[\int_0^t \frac{S(u\mid X)}{1-G(u-)}\lambda^*(u\mid X)du\right],
\end{align*}
where the first equality is due to the fact that $N$ can only jump once and thus $(\int_0^t dN(u)/(1-G(u-)))^2$ is simply equal to $\Delta I(T\leq t)/(1-G(T-))^2$.
Now,
\begin{align*}
\mathrm{MSE}'(t,\pi) &= \mathbb E\left[\left(1-\int_0^t \frac{dN(u)}{1-G(u-)}\right)^2\right]-2 \mathbb E\left[\left(1-\int_0^t \frac{dN(u)}{1-G(u-)}\right)\pi(t\mid X)\right]\\
&\quad+\mathbb E[(\pi(t\mid X))^2]\\
&=1-2\mathbb E[(1-S(t\mid X))]+\mathbb E\left[\left(\int_0^t \frac{dN(u)}{1-G(u-)}\right)^2\right]-2\mathbb E[S(t\mid X)\pi(t\mid X)]\\
&\quad+\mathbb E[(\pi(t\mid X))^2]\\
&=\mathrm{MSE^{Brier}}(t,\pi)+B(t),
\end{align*}
with
\begin{align*}
B(t)=-\mathbb E[1-S(t\mid X)]+\mathbb E\left[\int_0^t \frac{S(u\mid X)}{1-G(u-)}\lambda^*(u\mid X)du\right].
\end{align*}
Now, using Equation~\eqref{eq:Brier_eq}, we can rewrite $B(t)$ in the following way:
\begin{align*}
B(t)&=-\mathbb E\left[\int_0^t S(u\mid X)\lambda^*(u\mid X)du\right]+\mathbb E\left[\int_0^t \frac{S(u\mid X)}{1-G(u-)}\lambda^*(u\mid X)du\right]\\
&=\mathbb E\left[\int_0^t \frac{G(u-)}{1-G(u-)}S(u\mid X)\lambda^*(u\mid X)du\right].
\end{align*}
This shows that $B(t)\geq 0$ and that this quantity does not depend on $\pi$.

\subsection{Computation of $A(t)$ in the simulation section}\label{sec:A_comput}

While the inseparability term cannot be computed on real data, it is possible to obtain its expression when the distribution of all variables are known. In this section, we provide the explicit expression of the inseparability term $A(t)$, in the simulation setting of Section~\ref{sec:simu1}. To this end, we use Equation~\eqref{eq:A}. We first notice that, for $u<v$, $dN(u)dN(v)=I(C\geq v)dN^*(u)dN^*(v)$ and
\begin{align*}
\mathbb E[dN(u)dN(v)] & =\mathbb E[I(C\geq v)\mathbb E[dN^*(u)dN^*(v)\mid X,C]\\
 & = \mathbb E[I(C\geq v)\lambda_0(u)\lambda_0(v)\exp(2\theta_0^{\top} X)]dudv\\
 & = (1-G(v-))\lambda_0(u)\lambda_0(v)\mathbb E[\exp(2\theta_0^{\top} X)]dudv,
\end{align*}
where we used the fact that $\mathbb E[dN^*(u)dN^*(v)\mid X,C]=\mathbb E[dN^*(u)\mid X ] \mathbb E[dN^*(v)\mid X]=\lambda_0(u)\lambda_0(v) \exp(2\theta_0^{\top} X)dudv$, since under our simulation scheme, $N^*$ is independent of $C$ and $dN^*(u)$ is independent of $dN^*(v)$ conditionally on $X$, for $u\neq v$. Let $\gamma=3$, such that $C$ follows a uniform distribution on $[0,\gamma]$. For $t< \gamma$, we have:
\begin{align*}
\mathbb E\left[\left(\int_0^t\frac{dN(u)}{1-G(u-\mid X(u))}\right)^2\right]=A_1(t)+A_2(t),
\end{align*}
where
\begin{align}\label{eq:A_simu_proof}
A_1(t) & =2\iint_{0<u<v<t} \frac{\lambda_0(u)\lambda_0(v)}{1-G(u-)}dudv \mathbb E[\exp(2\theta_0^{\top} X)],\\
A_2(t) &= \int_0^t\frac{\mathbb E[dN(u)]}{(1-G(u-))^2}\cdot
\end{align}
We now compute
\begin{align*}
\int_{0}^v \frac{\lambda_0(u)}{1-G(u-)}du=2\times\frac{\gamma}{\beta^2} \int_{0}^v \frac{u}{\gamma-u}du=2\times\frac{\gamma}{\beta^2}\times\left(\gamma \log\left(\frac{\gamma}{\gamma-v}\right)-v\right) ,
\end{align*}
where we replaced $\lambda_0$ by the hazard of a Weibull distribution with shape parameter $\alpha=2$, scale parameter $\beta$ and the last equality was obtained from the change of variables $w=\gamma-u$. We then need to compute the following integral in $A_1(t)$:
\begin{align*}
 \int_0^t\left(\gamma \log\left(\frac{\gamma}{\gamma-v}\right)-v\right)\lambda_0(v)dv & = \frac{2}{\beta^2}\int_0^t\left(\gamma \log\left(\frac{\gamma}{\gamma-v}\right)-v\right)vdv\\
 &=  \frac{2}{\beta^2}\left(\frac{\gamma t^2}{2}\log(\gamma)-\gamma\int_0^t v \log(\gamma-v)dv-\frac{t^3}{3}\right).
\end{align*}
The last integral is computed using integration by parts and then by using the change of variables $w=\gamma-v$:
\begin{align*}
\int_0^t v \log(\gamma-v)dv & = \int_0^t \frac{v^2}{2} \frac{dv}{\gamma-v} + \frac{t^2}{2}\log(\gamma-t)\\
&=\int_{\gamma-t}^{\gamma}\frac{(\gamma-w)^2}{2w}dw + \frac{t^2}{2}\log(\gamma-t)\\
&=\frac{\gamma^2}{2}\log\left(\frac{\gamma}{\gamma-t}\right)-\gamma t+\frac{\gamma^2}{4}-\frac{(\gamma-t)^2}{4}+\frac{t^2}{2}\log(\gamma-t)\\
&=\frac{t^2-\gamma^2}{2}\log(\gamma-t)-\frac{\gamma t}{2}-\frac{t^2}{4}+\frac{\gamma^2}{2}\log(\gamma).
\end{align*}
Gathering all the parts in $A_1(t)$, we have:
\begin{align*}
& A_1(t)\\
&\quad = \frac{8\gamma}{\beta^4}\left(\frac{\gamma t^2}{2}\log(\gamma)-\frac{\gamma}{2}(t^2-\gamma^2)\log(\gamma-t)+\frac{\gamma^2 t}{2}+\frac{\gamma t^2}{4}-\frac{\gamma^3}{2}\log(\gamma)-\frac{t^3}{3}\right)\mathbb E[\exp(2\theta_0^{\top} X)].
\end{align*}
On the other hand, computation of $A_2(t)$ is straightforward, using the relation (see Section~\ref{sec:RC}) $\mathbb E[dN(t)\mid X(t)] =(1-G(t-)) \lambda^*(t\mid X(t))dt$. For $t<\gamma$, we have
\begin{align*}
A_2(t)&=\int_0^t \frac{\lambda_0(t)}{1-G(t-)} dt \mathbb E[\exp(\theta_0^{\top} X)]\\
 & = 2\times\frac{\gamma}{\beta^2}\times\left(\gamma \log\left(\frac{\gamma}{\gamma-t}\right)-t\right) \mathbb E[\exp(\theta_0^{\top} X)].
\end{align*}
Finally, according to Equation~\eqref{eq:A}, we need to compute $A_3(t)=\mathbb E\left[\left(\mu^*(t\mid X(t))\right)^2 \right]$. From Equation~\eqref{eq:simu_mu}, we directly have
\begin{align*}
A_3(t)= \left(\frac{t}{\beta}\right)^{2\alpha}\mathbb E[\exp(2\theta_0^{\top}X_i)].
\end{align*}
To conclude, $A(t)=A_1(t)+A_2(t)-A_3(t)$ and the terms involved in this equation including $\mathbb E[\exp(\theta_0^{\top}X_i)]$ or $\mathbb E[\exp(2\theta_0^{\top}X_i)]$ can easily be computed using Monte-Carlo simulations.

\bibliographystyle{biometrika}
\bibliography{biblio}

\end{document}